\documentclass[10pt,aps,prb,twocolumn,showpacs]{revtex4-1}
\usepackage{amsmath,amssymb,bm}
\usepackage{graphicx}

\begin{document}

\begin{abstract}
We study electronic states and topological invariants of (001)--films of topological crystalline insulator (TCI) Pb$_{x}$Sn$_{1-x}$Te.
Gapless surface Dirac cones on bulk TCIs become gapped in thin films due to  finite-size effect, which is hybridization between those on the top and bottom surfaces.
We clarify that the TCI film has the strong finite-size effect as compared to three-dimensional topological insulators such as Bi$_2$Se$_3$.
Moreover,
the energy gap oscillates with the thickness of film.
The oscillation stems from topological phase transitions in two dimensions.
The obtained data of the topological invariants and energy gap serve as guide to TCI-device applications.
\end{abstract}

\title{Finite-size-effect-induced topological phase transition in a topological crystalline insulator}

\author{Hideyuki Ozawa} 
\affiliation{Department of Applied Physics, Nagoya University, Nagoya 464-8603, Japan}
\author{Ai Yamakage}
\affiliation{Department of Applied Physics, Nagoya University, Nagoya 464-8603, Japan}
\author{Masatoshi Sato}
\affiliation{Department of Applied Physics, Nagoya University, Nagoya 464-8603, Japan}
\author{Yukio Tanaka}
\affiliation{Department of Applied Physics, Nagoya University, Nagoya 464-8603, Japan}

\date{\today}
\maketitle

\section{Introduction}

Topological insulators (TIs) are a new state of matter which supports Dirac fermions on its surface and exhibit novel phenomena resulting from the Dirac fermions.\cite{HasanKane, qi11, ando13}
A prototypical TI is quantum Hall insulator in which time-reversal symmetry is broken.
In recent years, on the other hand,
time-reversal symmetric TI has attracted much attention.
Interestingly, symmetry, e.g., time-reversal, gives rise to nontrivial topological number and related topological phenomena.
Various symmetries in condensed matters are expected to yield diverse topological materials.

Surface Dirac fermions on TIs are protected by time-reversal symmetry.
Nowadays, so-called topological crystalline insulators (TCIs),\cite{fu11} which are a nontrivial insulator supporting surface Dirac fermions protected by crystal symmetry,  has been proposed\cite{hsieh12} and experimentally discovered\cite{tanaka12, xu12, dziawa12} in IV--VI semiconductors.
Topological protection by crystalline symmetry enables us to find new topological systems even in insulators which have been thought to be topologically trivial.
Moreover, the mechanism different from that of TI can lead to different topological phenomena.
Indeed,
many materials \cite{wojek13, tanaka13, gyenis13, okada13, safaei13} and theoretical studies \cite{slager13, kargarian13, wang13, ye13, barone13, liu13, weng13, sun13, zhang13, liuXJ13, yokoyama14} on TCI have been reported.
Furthermore, superconductivity\cite{erickson09} and its topological non-triviality\cite{sasaki12} in a doped TCI has been observed and attracted attentions.\cite{balakrishnan13, zhong13, sato13, he13, goswami13, novak13, yu14}

TCIs are not so robust since the crystal symmetry can be easily broken by an external field.
Nonetheless, this behavior can be applied to a highly controllable device: an external electric field breaking the crystal symmetry may control the gap of the surface Dirac fermions.
Note that this is a great advantage for device application. 
TI device, in principle, can be realized but needs a magnetic field, which is not convenient in a nano-sized system, to open and control energy gap in the surface Dirac fermions.
An electric field is, on the other hand, easily implemented in devices, such as field-effect transistor.

From the perspective of topological material design and its application, nano-fabrication, e.g., thin film and heterostructure, is one of the most promising ways:
the number of careers in thin films can be highly controlled by applying gate voltage, and
electronic states and its dimensionality can be dramatically tuned in heterostructures.
Indeed, there are many studies  on thin films\cite{linder09, lu10, liu10,  shan12, ebihara12, imura12, singh13, okamoto14, bjyang14, sakamoto10, zhang10, taskin12} and heterostructures\cite{burkov11, fukui13, simpson11, tominaga11, nakayama12, lin13} of TIs. 
Furthermore,
several studies on nano-fabrication of TCIs have been reported;
field-effect devices with using TCI,\cite{fang14, liu14, ezawa14}
 TCI heterostructure,\cite{yang14}
 and
experimental fabrication of TCI films.\cite{taskin13, yan14, assaf14}
Now device application of TCIs is becoming active.

In this paper, we study electronic states and topological invariants of thin film of Pb$_x$Sn$_{1-x}$Te.
In the thin film, 
the surface Dirac cones become gapped since the wave functions of the top and bottom surface states hybridize with each other.
As the number of layers $N_z$ increases,
the induced gap decays exponentially but slowly as compared to TIs such as Bi$_2$Se$_3$.
For the odd numbers of layers, we find that the energy gap shows a damped oscillation as a function of $N_z$, which stems from two types of topological phase transition in two dimensions: one is that between a trivial to two-dimensional TCIs, and the other is that between two TCIs.
The former transition is in agreement with that obtained in Ref. \onlinecite{liu14}.
And also,  a similar damped oscillation of the energy gap is found in the even numbers of layers, which accompanies another topological phase transition.
These non-monotonic change of the energy gap and topological phase are qualitatively and quantitatively clarified. 
Our results are useful for experiments on thin films of TCIs.
 
 The paper is organized as follows.
In Sec. \ref{secgap}, we present electronic states of a (001)--film of Pb$_x$Sn$_{1-x}$Te.
Thickness dependence of the energy gap is closely shown.
The causes of this dependence are resolved in terms of topological invariants in Sec. \ref{sectp}.
Gapless edge states related to the topological invariants are also clarified.
Finally we summarize our results in Sec. \ref{summary}.
The details of the model Hamiltonian, symmetry, and the topological numbers are explained in Appendices.

\section{Energy gap induced by finite-size effect}
\label{secgap}

Bulk Pb$_x$Sn$_{1-x}$Te is a TCI, which supports gapless Dirac cones protected by the (110)--reflection symmetry.
Firstly, we review the gapless surface states on the bulk system.
Next, we show extensive data on the energy gap induced by the finite-size effect in thin films of Pb$_x$Sn$_{1-x}$Te.

\subsection{Gapless surface states on the bulk}

\begin{figure}
\centering
\includegraphics{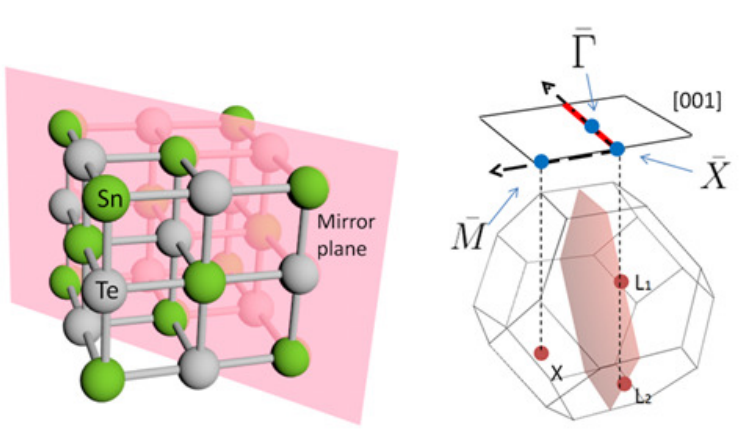}
\caption{Crystal structure of SnTe, the first Brillouin zone, and its projection onto the (001) plane.
Mirror symmetry with respect to (110) plane illustrated in the figure protects the surface Dirac cones.
}
\label{crystal}
\end{figure}

We use a model Hamiltonian of IV--VI semiconductors with $s$, $p$, and $d$ orbitals.\cite{lent86}
The fcc crystal structure and the corresponding first Brillouin zone are illustrated in Fig. \ref{crystal}.
\begin{figure}
\centering
\includegraphics{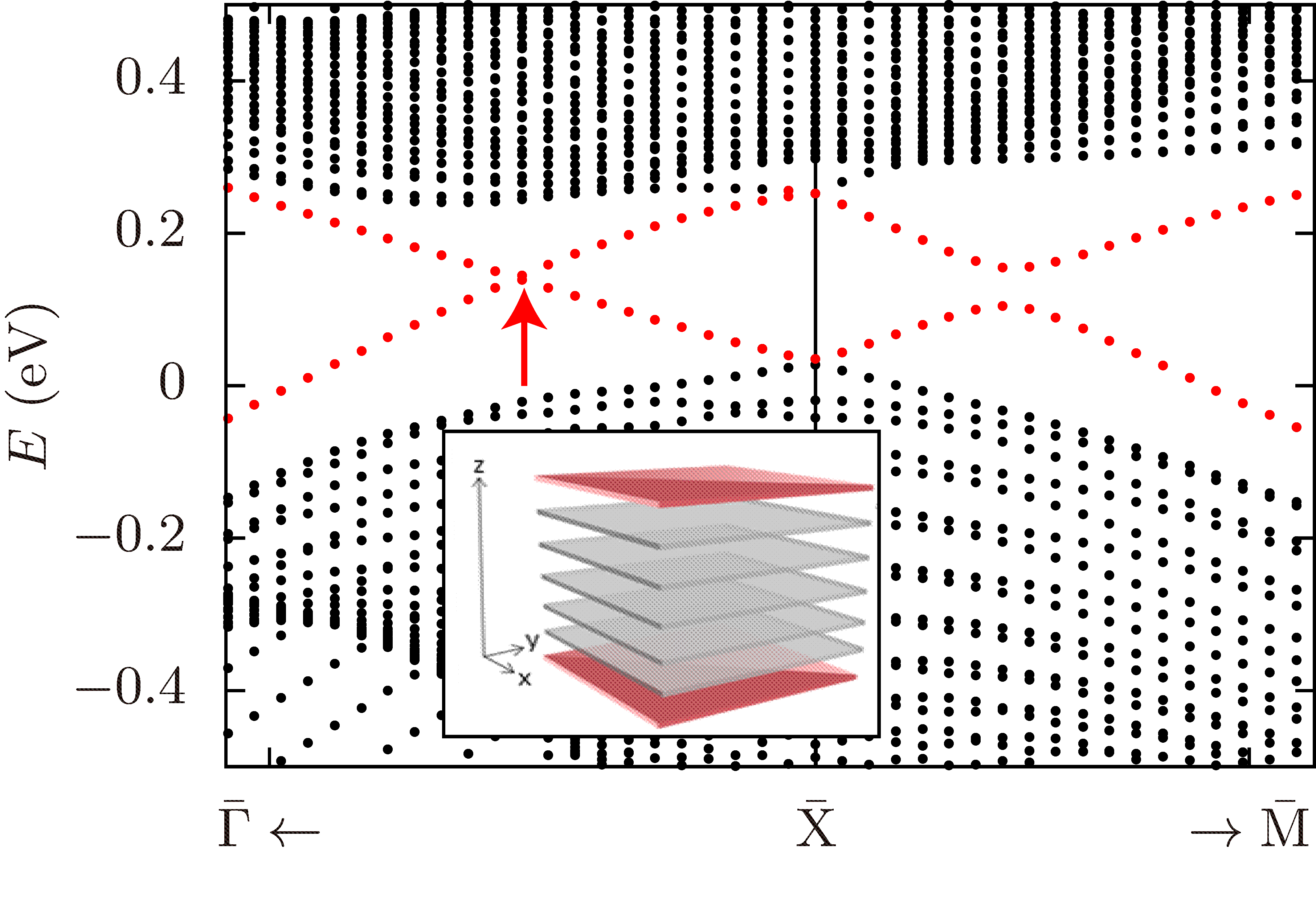}
\caption{Energy dispersion for SnTe near the $\rm \bar X$ point in slab geometry shown in the inset.
The arrow denotes the Dirac point of the surface states.
The number of layers is set to $N_z= 45$. 
The parameters for SnTe are taken from Ref. \onlinecite{lent86}.}
\label{surface}
\end{figure}
The bottom of conduction band and top of valence band are located near the L points.
The L points are projected onto the $\rm \bar X$ points on the (001) surface. 
Figure \ref{surface} shows an energy dispersion for a thick SnTe slab, where
two surface Dirac cones are located on the $\rm \bar \Gamma \bar X$ line.
The model of IV--VI semiconductor with the (001) surface is explicitly shown in Appendix \ref{model}.
The Dirac cones are protected by the mirror Chern number in the (110) surface.\cite{hsieh12}
On the contrary, there is no gapless surface states along the $\rm \bar X \bar M$ nor $\rm \bar \Gamma \bar M$ lines.
Note that in this model the origin of energy is set to the top of valence bands,
and
the Fermi energy is located at the Dirac point ($E \sim 0.15$eV) for charge neutral cases.

\subsection{Strong finite-size effect in SnTe}

\begin{figure}
\centering
\includegraphics{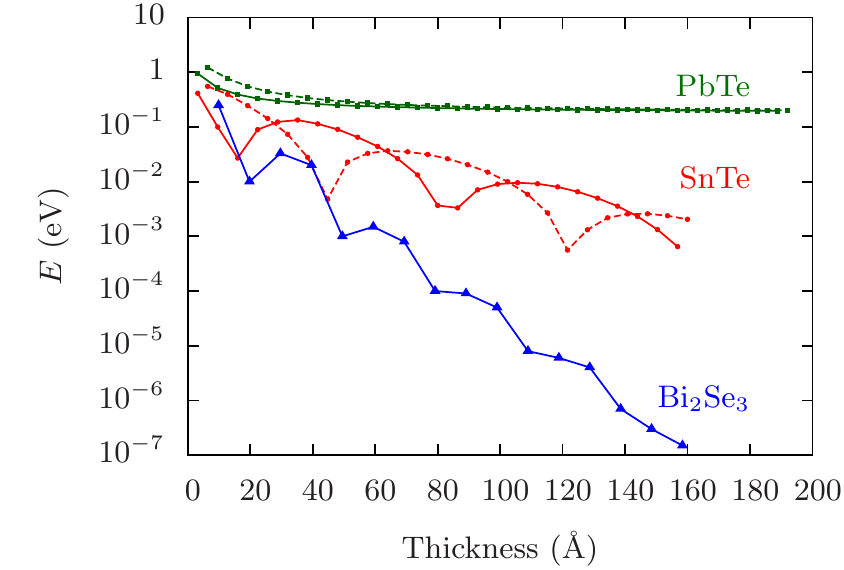}
 \caption{Finite-size-effect-induced energy gap for thin films of non-topological PbTe [(green) square], TCI SnTe [(red) circle], and TI Bi$_2$Se$_3$ [(blue) triangle]  as a function of the thickness.
The solid and dashed lines denote the magnitude of energy gap for the odd and even numbers of layers, respectively.}
\label{gap1}
\end{figure}

Next we turn to energy gap in the (001) thin film.
A wave function of the gapless surface states decays exponentially into the bulk $\psi \sim e^{-z/\xi}$ with the penetration depth $\xi$.
For the thin film case, 
the tails of wave functions of the surface states localized on the top and bottom surfaces overlap with each other.
In consequence, energy gap ($\sim e^{-N_z/\xi}$) of the surface states is induced due to the finite-size effect.
The induced gap for TCI SnTe film are shown in Fig. \ref{gap1}.
The induced gap for TI Bi$_2$Se$_3$ film (see Appendix \ref{bi2se3}), and energy gap for a trivial insulator PbTe film are also shown as a reference.
The energy gap of PbTe film is nearly independent of the thickness, and reaches to be 0.2eV which is the magnitude of the bulk gap, since PbTe has no gapless surface state.
On the other hand, the energy gap of SnTe film decreases exponentially ($\xi \sim 10$ layers $\sim 30$ \AA) and oscillates as the thickness increases.
We firstly focus on the damping part. The oscillation part will be discussed in the next section.
In the thick limit, the energy gap of SnTe vanishes and gapless surface Dirac cones are reproduced on the (001) surface.
As a reference, the energy gap of Bi$_2$Se$_3$ film is shown.
The energy gap decreases and oscillates more quickly ($\xi \sim 1$ quintuple layer $\sim 10$\AA) than that of SnTe film.
The difference comes from crystal structure: 
SnTe is a cubic crystal, and Bi$_2$Se$_3$ is a rhombohedral crystal with quintuple layered structure. 
Namely, in SnTe,
all the bondings  are equivalent. 
But in Bi$_2$Se$_3$, the bonding between the quintuple layers is weak.
This is the reason why the finite-size effect in (001) thin film of SnTe is much stronger than that of Bi$_2$Se$_3$.

\begin{figure}
\centering
\includegraphics{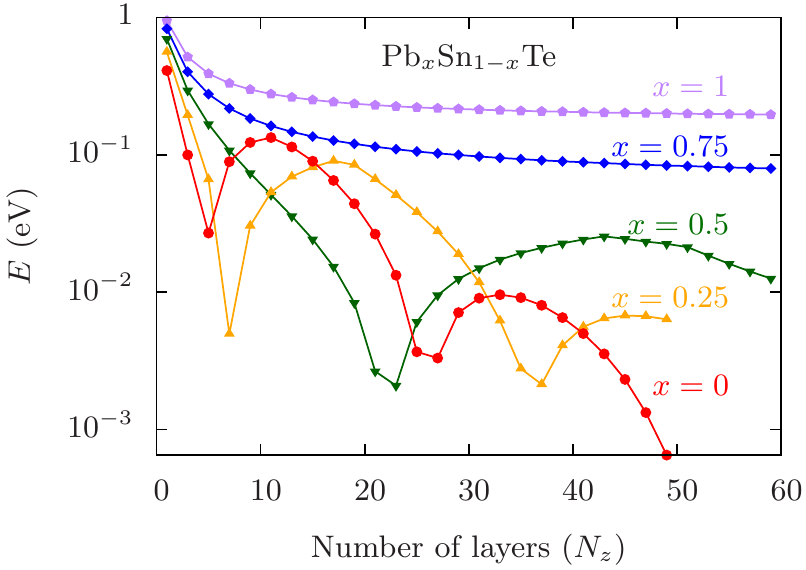}
\caption{Energy gap in thin film of Pb$_x$Sn$_{1-x}$Te for various compositions $x$ for the odd numbers of layers.}
\label{gap2}
\end{figure}
We quantitatively clarify the magnitude of the finite-size-effect induced gap in Pb$_x$Sn$_{1-x}$Te thin films (Fig. \ref{gap2}).
For $x \geq 0.75$, the system is in the trivial phase. As the thickness increases, the energy gap decreases exponentially and converges to be the bulk band gap (0.2eV for PbTe and 0.08eV for Pb$_{0.75}$Sn$_{0.25}$Te) in the thick limit.
At $0.5 < x < 0.75$,
the topological phase transition occurs and the system falls into the TCI phase in $x \leq 0.5$.
In the TCI phase,
the energy gap decays as $\sim e^{-N_z/\xi}$.
%
As can be seen from Fig. \ref{gap2}, 
the decay ratio of the energy gap ($\sim \xi^{-1}$) is proportional to
the \textit{distance} from the topological phase transition point.
Just at the transition point ($0.5 < x < 0.75$), the energy gap decays extremely slowly ($\xi \to \infty$).

\subsection{Oscillation of the energy gap}

The calculated results shown in Figs. \ref{gap1} and \ref{gap2} exhibit oscillation of the energy gap for the TCI films in addition to the exponential decay as a function of the thickness.
Such a damped oscillation is known to exist in TI Bi$_2$Se$_3$.\cite{linder09, lu10, liu10, shan12, ebihara12, imura12, singh13, okamoto14}
Moreover, an even-odd effect also appears in the energy gap (see  SnTe in Fig. \ref{gap1}).
This behavior is understood intuitively as follows: the top and bottom layers are the same (different) for odd (even) numbers of layers of the (001) film.
Actually, the even and odd numbers of layers have different symmetries and are characterized by different topological invariants, as we shall see in Sec. \ref{sectp}.

\begin{figure}
\centering
\includegraphics{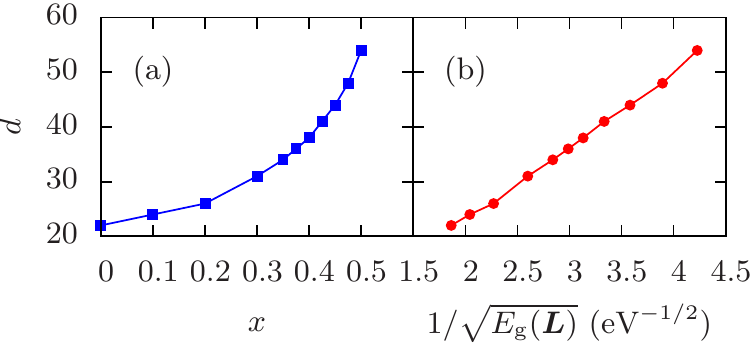}
\caption{Period $d$ of oscillation of the energy gap in the film in unit of the number of layers as a function of $x$ (a) and of $1/\sqrt{E_{\rm g}(\bm L)}$ (b)
 with $E_{\rm g}(\bm L)$  the bulk band gap at the L point.}
 \label{period-gap}
\end{figure}

Here we show the details on the oscillation for odd numbers of layers of Pb$_x$Sn$_{1-x}$Te (Fig. \ref{gap2}).
In the trivial phase ($x \geq 0.75$) the energy gap does not oscillate.
At the topological phase transition point (located at $0.5 < x < 0.75$), the energy gap closes in $N_z \to \infty$, i.e., the period of the oscillation is regarded to be infinity.
Apart from the transition point, $x \leq 0.5$, the period becomes shorter, and takes to be about 20 layers for $x=0$.
On the other hand, the period of the oscillation in Bi$_2$Se$_3$ is about three quintuple-layers, which is one order of magnitude shorter than that of SnTe.
Namely, in materials with the strong finite-size effect, the magnitude of the energy gap slowly decays and oscillates with a long period as a function of the thickness.
Note that the period in TI films is given by $\pi \sqrt{B/M_0}$,\cite{liu10} where $M_0$ and $B$ correspond to the magnitude of bulk band gap and effective mass, respectively (see Appendix \ref{bi2se3}).
We find an empirical rule for the period in a Pb$_x$Sn$_{1-x}$Te film, which is similar to that for TI films.
The period is scaled by $\sqrt{1/E_{\rm g}(\bm L)}$ with $E_{\rm g}(\bm L)$ the magnitude of band gap at the L point in the bulk, as shown in Fig. \ref{period-gap}.

\subsection{Discussion}

To summarize the finite-size-effect-induced energy gap of TCI films, 
a damped oscillation occurs in the energy gap as a function of the thickness, whose period is about 20 layers ($\sim 60$\AA).
A TCI film has an advantage over the bulk system: bipolar transport can be realized with applying a gate voltage.
The Fermi level can be tuned also to the Dirac point on the surface states of the bulk TCI.
Our result shows the penetration depth of wave function of the surface states in the film is about 10 layers ($\sim 30$\AA).
In order to use Pb$_x$Sn$_{1-x}$Te film as a three-dimensional TCI,
the number of layers must be set to $N_z \gg 10$.
Alternatively, it is useful that the thickness is set to the \textit{hot spot} (for SnTe, $N_z \sim 20$\AA, 40\AA, 80\AA, 120\AA, $\cdots$, see Fig. \ref{gap1}) which is a minimal value in the damped oscillation.

With decreasing the thickness, the system crossovers from three-dimensional to two-dimensional insulators around $N_z \sim \xi \sim 10$, energy gap of the surface states becomes larger.
The crossover is schematically summarized in Fig. \ref{crossover1}.
In exchange for gapless surface states, the thin film works as a two-dimensional TCI with one-dimensional edge states, which will be explained in the next session.

\section{Topological phase diagram}
\label{sectp}

In this section, we show that
oscillatory behavior of the energy gap found in the previous section stems from  topological phase transitions, 
i.e., the energy gap takes a minimal value in the vicinity of the transition point.
The corresponding gapless edge states are also discussed.
%
The film has the (001)--reflection symmetry defined by 
$\mathcal M H(k_x, k_y) \mathcal M^\dag = H(k_x, k_y)$ with
\begin{align}
\label{def1} 
\mathcal M c_{n}(k_x, k_y) \mathcal M^\dag = \eta M c_{N_z+1-n}(k_x, k_y),
\
M = -i P_z s_z, 
\end{align}
where $H$ is Hamiltonian of the film, $P_z$ is the (001)--reflection operator acting on the orbitals, $s_z$ is the $z$-component of spin, $c_n$ is the annihilation operator in the $n$-th layer for $n=1, \cdots, N_z$, and $\eta$ is a phase factor as $\eta=-1$ for $N_z=4m-3$, $N_z=4m-2$, 
and $\eta=+1$ for $N_z=4m-1$, $N_z=4m$ with $m \in \mathbb N$.
Since the (001)-reflection symmetry is preserved in the odd numbers of layers,\cite{liu14}  one can define the mirror Chern number.
Additionally,
the (001)--reflection symmetry can be extended to the even numbers of layers, as explained in Appendix \ref{symmetry}.
As a result, the Hamiltonian of the even and odd numbers of layers is decomposed into the mirror-even $H_+$ and mirror-odd $H_-$ sectors, which has the definite mirror eigenvalue of $M=+i$ (mirror-even) and $M=-i$ (mirror-odd),  as
\begin{align}
 H(k_x, k_y) &= H_+(k_x, k_y) \oplus H_-(k_x, k_y),
 \\
 H_\pm(k_x, k_y) &= P_\pm H(k_x, k_y) P_\pm,
\end{align}
with the projection operator $P_\pm$ onto the mirror-even/odd sector
\begin{align}
P_\pm = \sum_m | \pm, m \rangle \langle \pm, m|,
\
\mathcal M |\pm, m \rangle = \pm i | \pm, m \rangle,
\end{align}
with $|\pm , m\rangle$ being the $m$--th eigenvector of $\mathcal M$ with an eigenvalue of $\pm i$.

\begin{figure*}
\centering
\includegraphics[scale=0.7]{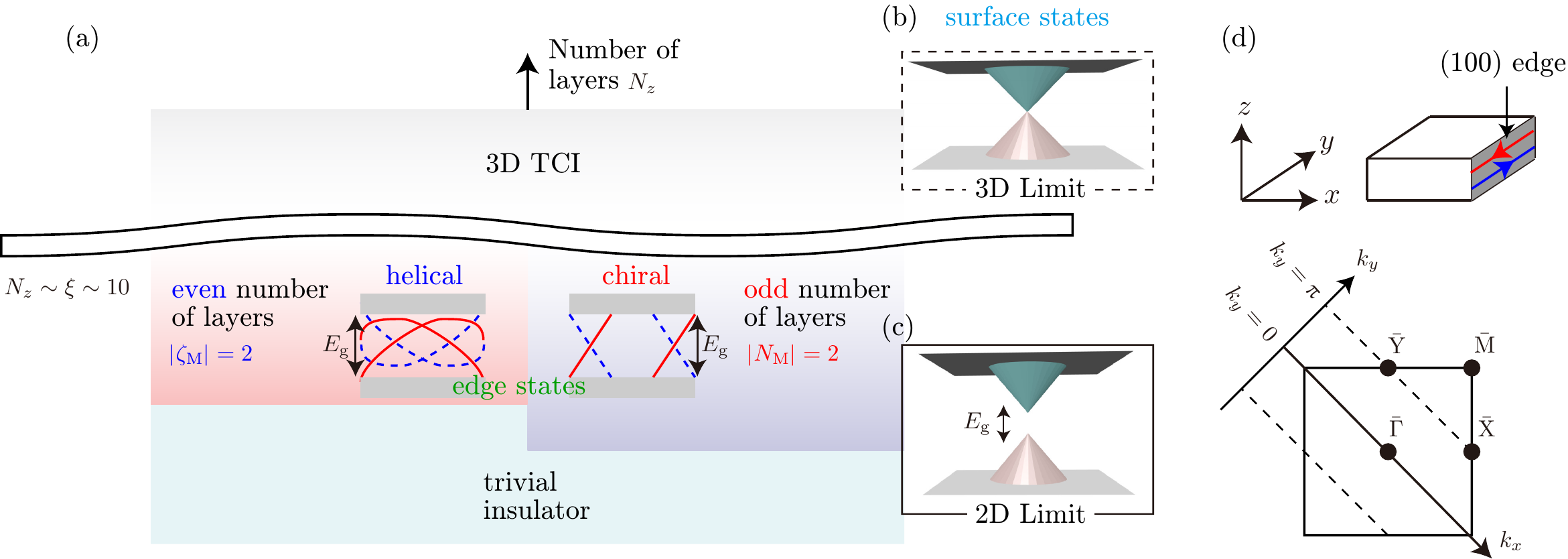}
\caption{
Dimensional crossover (a) from three-dimensional (3D) to two-dimensional (2D) insulators in a Pb$_x$Sn$_{1-x}$Te film.
The two-dimensional gapless surface states (b) has a large gap in thin film of $N_z < \xi \sim 10$ (c).
The mirror-Chern phase $|N_{\rm M}|=2$ supporting the mirror-chiral gapless edge states and the $|\zeta_{\rm M}|=2$ phase, which is defined by Eq. (\ref{zeta}), supporting the mirror-helical gapless/gapful edge states, are realized for the odd and even numbers of layers, respectively.
Energy dispersions of the edge states are schematically shown.
The (red) solid and (blue) dashed line denote those of edge states in the mirror-even and mirror-odd sectors.
In the thin limit, the system falls into a trivial insulator.
The two dimensional and the projected one-dimensional BZs are illustrated in (d).
}
\label{crossover1}
\end{figure*}

Differently from the odd numbers of layers,
since time-reversal symmetry is preserved in each mirror sector for the even numbers of layers (see Appendix \ref{trs}),
the mirror Chern number vanishes.
Instead, we introduce a topological invariant characterizing bulk energy bands and gap in the even numbers of layers.
The results for the even and odd numbers of layers are summarized in Fig. \ref{crossover1}.
The details are explained in the following sections.

\subsection{Odd numbers of layers}

\subsubsection{Phase diagram}

\begin{figure}
\centering
\includegraphics{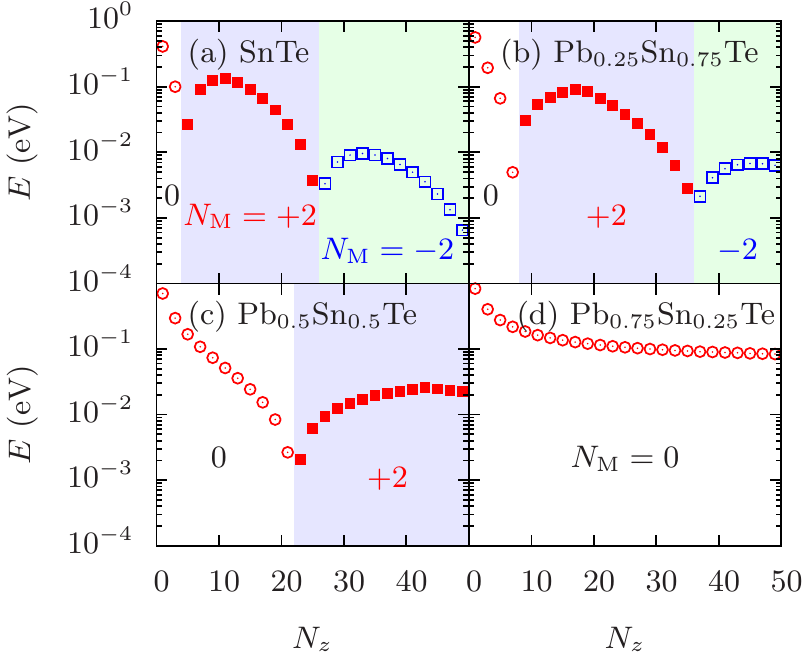}
\caption{Topological phase diagram in odd numbers of layers of Pb$_x$Sn$_{1-x}$Te.
TCI with $|N_{\rm M}|=2$ is realized in the shaded regions.
Energy gap is denoted by the open circles, closed squares, and open squares for non-topological insulator $N_{\rm M}=0$, TCI with $N_{\rm M}=2$, and TCI with $N_{\rm M}=-2$, respectively.
}
\label{pd_odd}
\end{figure}

The topological phase of the odd numbers of layers is characterized by the (001)--mirror Chern number  $N_{\rm M}$,\cite{liu14}
 which is defined by
\begin{align}
 N_{\rm M} = \frac{N_+ - N_-}{2},
\end{align}
where $N_+$ and $N_-$ denote the Chern number for the mirror-even and mirror-odd sectors, respectively.
$N_\pm$ is calculated by the method proposed in Ref. \onlinecite{fukui05}.
The obtained topological phase diagram for the odd numbers of layers is shown in Fig. \ref{pd_odd} with the magnitude of the energy gap.
For the odd numbers of layers of SnTe film with $N_z \leq 3$ and $N_z \geq 5$, the mirror Chern number $N_{\rm M}$ is obtained to be $N_{\rm M} = 0$  and $|N_{\rm M}|=2$, respectively.
The topological phase transition occurs between $N_z=3$ and $N_z=5$.
At this time the band gap closes at the $\rm \bar X$ point.\cite{liu14}
This is why the energy gap takes a minimal value between $N_z=3$ and $N_z=5$.
The same tendency is seen for $x=0.25$ and $x=0.5$: the topological phase transition from trivial to $|N_{\rm M}|=2$ insulators occurs at $7<N_z<9$ for $x=0.25$ and at $21 < N_z < 23$ for $x=0.5$.
On the other hand, for $x \geq 0.75$, there exists no non-trivial phase for any $N_z$.

\subsubsection{Edge state}

Next we investigate detail electronic states of the (100)--edge states.
The one-dimensional projected Brillouin zone onto the (100) edge is defined in $k_y \in [-\pi, \pi]$ as illustrated in Fig. \ref{crossover1}(d).
Edge spectral function $\rho_\pm(k_y, E)$ in each mirror sector is defined by
\begin{align}
 \rho_\pm(k_y, E) = \mathrm{Im} \lim_{N_x \to \infty}\left. \frac{1}{E-\tilde H_\pm(k_y) + i0} \right|_{n_x=N_x},
\end{align}
in the semi-infinite system,\cite{umerski} where $\tilde H_\pm(k_y) = P_\pm \tilde H(k_y) P_\pm$ is the Hamiltonian in each mirror sector  and  $\tilde H(k_y)$ is a Hamiltonian of the film with (100) defined in Appendix \ref{edgemodel}.
The edge charge $\rho_{\pm, \rm c}(k_y, E)$ and spin $\rho_{\pm, \rm s}(k_y, E)$ spectral functions for the mirror-even and mirror-odd sectors are defined by
\begin{align}
 \rho_{\pm, \rm c}(k_y, E) &= \frac{1}{18N_z} \mathrm{Tr} \left[ \rho_\pm(k_y, E) \right],
 \\
 \rho_{\pm, \rm s}(k_y, E) &= \frac{1}{18N_z}\mathrm{Tr} \left[ s_z \rho_\pm(k_y, E)\right],
\end{align}
where the normalization factor of $1/18$ is multiplied since each mirror sector has 18 bands.
The spin and mirror resolved edge density of states (DOS) $\rho_{\pm, s}$ is given by
\begin{align}
 \rho_{\pm, \uparrow}(E) &= 
\int \frac{dk_y}{2\pi}
\frac{\rho_{\rm c}(k_y, E) + \rho_{\rm s}(k_y, E)}{2},
\\
 \rho_{\pm, \downarrow}(E) &= 
\int \frac{dk_y}{2\pi}
\frac{\rho_{\rm c}(k_y, E) - \rho_{\rm s}(k_y, E)}{2}.
\end{align}

\begin{figure*}
\centering
\includegraphics{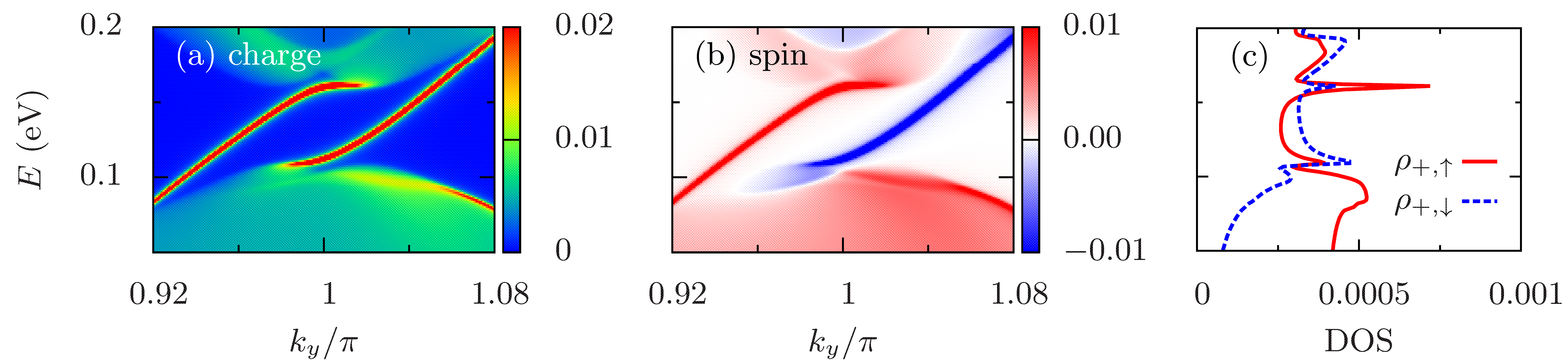}
\caption{
The edge charge $\rho_{+, \rm c}(k_y, E)$ (a) and spin $\rho_{+, \rm s}(k_y, E)$ (b) spectral functions and the spin-resolved edge DOS $\rho_{+, \uparrow}(E)$ and $\rho_{+, \downarrow}(E)$ (c) for the mirror-even sector.
The thickness is set to $N_z=7$.
All the vertical axes denote energy $E$ (eV) in the same scale.
}
\label{spin}
\end{figure*}

The calculated results for the mirror-even sector are shown in Fig. \ref{spin}.
The thickness is set to $N_z=7$, where the mirror Chern number is obtained to be $N_{\rm M} = 2$.
The charge spectral function [Fig. \ref{spin}(a)] clearly show two branches of gapless mirror-chiral edge states, 
which are protected by the mirror Chern number $N_{\rm M} = 2$.
As the edge states have been expected to be spin-filtered,\cite{liu14}
we evaluate the spin spectral function [Fig. \ref{spin}(b)]. 
This indicates that
the gapless edge states in the left and right branches are composed of spin up and down, respectively. 
Both edge states with spin up and down can go to the same direction, i.e.,
the spin of edge states is not completely but partially filtered.
In the edge DOS, spin polarization is estimated to be
$(\rho_{+ , \uparrow}-\rho_{+, \downarrow})/ (\rho_{+ , \uparrow} + \rho_{\downarrow}) \sim 0.1$ within the bulk gap
as shown in Fig. \ref{spin}(c).

 In the present system, the (001)--reflection operator $\mathcal M$ is given by $\mathcal M c_{n} \mathcal M^\dag = \eta M c_{N_z+1-n}$ and $M = -i P_z s_z$ [see Eq. (\ref{def1})], which depends not only on spin but also on orbitals and layer: spins of $p_x$-- and $p_y$--orbitals are canceled with that of $p_z$--orbital in the (001)--reflection, and the reflection is nonlocal with respect to degrees of freedom of layer $n$.
As a consequence, the (001)--reflection operator is no longer regarded as the well-defined spin.
But the partially polarized $z$-component of spin [Fig. \ref{spin}(c)] may give rise to spin transport phenomena such as spin Hall conductance.
For realization of a spintronics device with TCI films, more extensive studies on the edge states is needed.

\begin{figure*}
\centering
\includegraphics{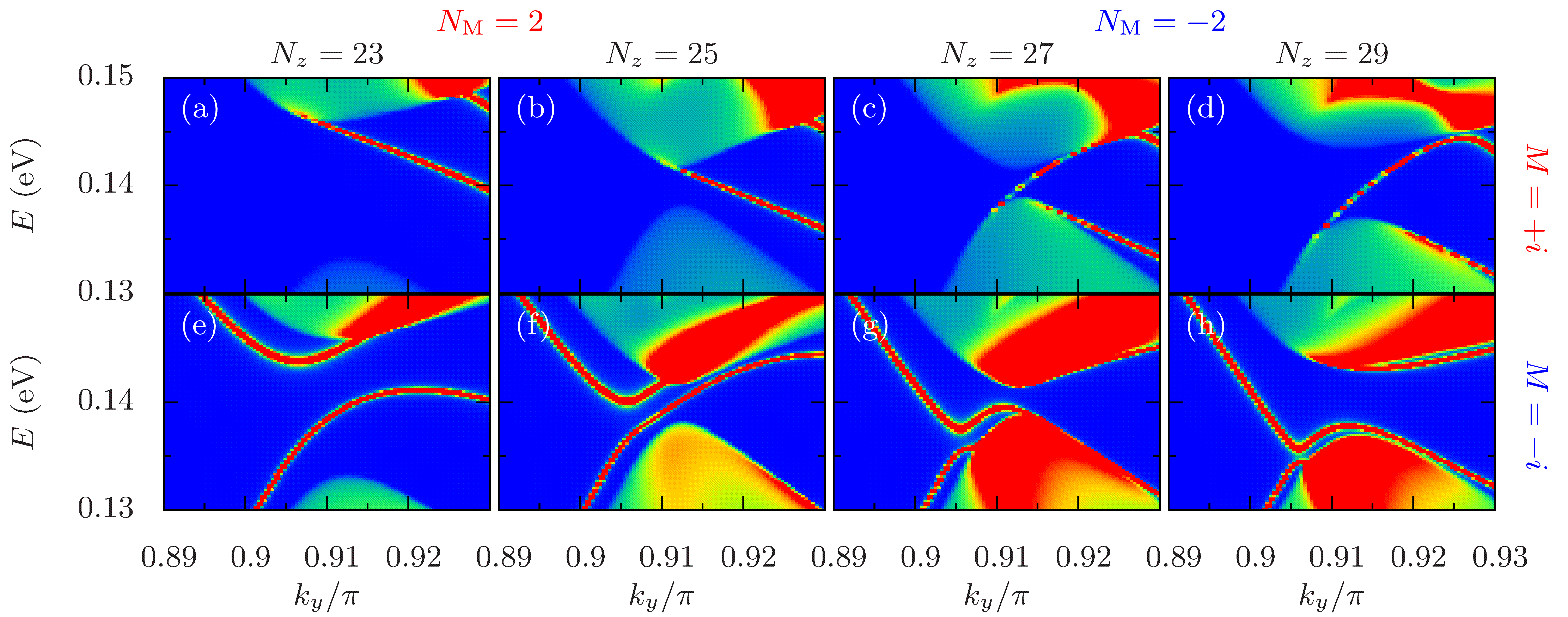}
\caption{
Edge spectral function $\rho_{\pm, \rm c}(k_y, E)$ for $N_z=23$ [(a) and (e)], 25 [(b) and (f)], 27 [(c) and (g)], and 29 [(d) and (h)].
The mirror eigenvalue is given by $M=+i$ for (a)--(d) and $M=-i$ for (e)--(h).
The mirror Chern number is obtained to be $N_{\rm M}=2$ for $N_z \leq 25$ and $N_{\rm M}=-2$ for $N_z \geq 27$.
}
\label{23-29}
\end{figure*}

The second minima in the energy gap appear at $N_z = 25$ for SnTe [Fig. \ref{pd_odd}(a)] and at $N_z = 35$ for Pb$_{0.25}$Sn$_{0.75}$Te [Fig. \ref{pd_odd}(b)].
This stems from the sign-change of the mirror Chern number at these minima, where the band gap closes simultaneously at two momenta away from the $\rm \bar X$ point.
Note that there is an ambiguity on the sign of the mirror Chern number since the phase factor of the reflection operator $\mathcal M$ is arbitrary.
The phase is fixed to $\eta$ in Eq. (\ref{def1}).
Validity of this choice is proven by confirming the detail structure of energy dispersion for the edge states shown in Fig. \ref{23-29}.
%
The finite--size--effect induced gap is located at $k_y \sim 0.91 \pi$, and there exists gapless edge states within the gap.
In the $N_{\rm M}=2$ phase, the group velocity of the edge state in the $M=+i$ sector is negative [Fig. \ref{23-29}(a) and \ref{23-29}(b)] near $k_y \sim 0.91 \pi$.
The bulk gap shows a minimum at $25 < N_z < 27$ then
the group velocity becomes positive in the $N_{\rm M}=-2$ phase [Fig. \ref{23-29}(c) and \ref{23-29}(d)].
The sign change of the mirror Chern number captures change of the local structure of energy dispersions for the edge states, i.e., the sign of the mirror Chern number corresponds to that of the edge state in the vicinity of the bulk band gap.
And also, the similar situation occurs in the $M=-i$ sector [Fig. \ref{23-29}(e)--\ref{23-29}(h)].
The above discussion justifies the definition Eq. (\ref{def1}) and the choice of the phase factor $\eta$.

\subsection{Even numbers of layers}
\subsubsection{Phase diagram}
The mirror Chern number $N_{\rm M}$ vanishes in the even numbers of layers since time-reversal symmetry [see Eq. (\ref{trs2})] is preserved.
In stead of the mirror Chern number, 
a winding number $\zeta_{\rm M}$ becomes nontrivial.
$\zeta_{\rm M}$ is defined by $\zeta_{\rm M} = (\zeta_+ - \zeta_-)/2$ with
\begin{align}
\label{zeta}
 \zeta_{\pm} = 
  \int_{0}^{2\pi} \frac{dk_y}{2\pi} \frac{\partial \theta_\pm(k_y)}{\partial k_y}
 - \frac{  \theta_\pm( {2\pi} )  - \theta_\pm(0)}{2\pi},
\end{align}
and
\begin{align}
 \theta_\pm(k_y)  
&= \mathrm{Im} \ln \det 
 \biggl\{
\mathrm P 
\exp \left[
  i \int_{-\pi}^{\pi} dk_x A_\pm(k_x,k_y)
 \right] 
 \nonumber\\ & \hspace{6em} \times
 B_\pm(k_y)
\biggr\},
\end{align}
where $\mathrm P$ stands for the path-ordering.
The non-Abelian Berry connection $A_\pm$ is defined by\cite{berry84, wilczek84}
\begin{align}
 [A_\pm(k_x, k_y)]_{mn} = \langle k_x, k_y, \pm, m | \frac{\partial}{\partial k_x}
 | k_x, k_y, \pm, n \rangle,
\end{align}
where $| k_x, k_y, \pm, m \rangle$ is an eigenvector of the Hamiltonian $H_\pm(k_x, k_y)$, and $m$, $n$ are the occupied band indices.
Matrix $B(k_y)$, which is defined by
\begin{align}
 [B_\pm(k_y)]_{mn} = 
\langle \pi, k_y, \pm, m 
\left|  -\pi, k_y, \pm, n 
\right\rangle,
\end{align}
is attached in Eq. (\ref{zeta}) so that $\zeta_{\rm M}$ is invariant for gauge transformation and symmetry operations of the system.
$\zeta_{\rm M}$ introduced above is a $\mathbb Z$ topological invariant in two spatial dimension.
This invariant is protected by translation, reflection, and time-reversal symmetries, which is explained in Appendix \ref{mz2}.
A numerical recipe for $\zeta_{\rm M}$ is given also in Appendix \ref{mz2}.

\begin{figure}
\centering
\includegraphics{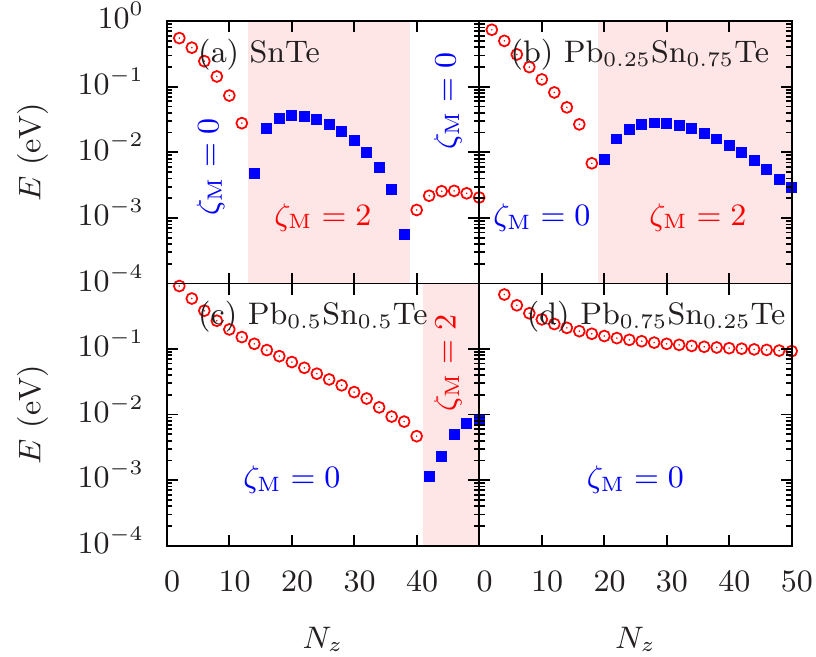}
\caption{
Topological phase diagram in even numbers of layers of Pb$_x$Sn$_{1-x}$Te.
Energy gap and topological invariant $\zeta_{\rm M}$ are shown. 
$|\zeta_{\rm M}|=2$ is denoted by the
closed square in the shaded region.}
\label{pd_even}
\end{figure}

The calculated topological invariant are shown in Fig. \ref{pd_even} with the energy gap.
In the thin limit, the topological number is obtained to be $\zeta_{\rm M}=0$.
As the thickness increases, the energy gap takes a minimal value at $12< N_z < 14$ for SnTe,  at $18 < N_z < 20$ for Pb$_{0.25}$Sn$_{0.75}$, and at $40 < N_z < 42$ for Pb$_{0.5}$Sn$_{0.5}$Te, when the topological number changes from $\zeta_{\rm M}=0$ to $|\zeta_{\rm M}|=2$.
The change of topological invariant must be associated by bulk gap closing.
This is why the energy gap exhibits minima  as the thickness increases in the even numbers of layers.

\subsubsection{Edge state}
Here we discuss gapless one-dimensional edge states on the (100) edge [see Fig. \ref{crossover1}(d)] in the case of $|\zeta_{\rm M}|=2$.
\begin{figure}
\centering
\includegraphics{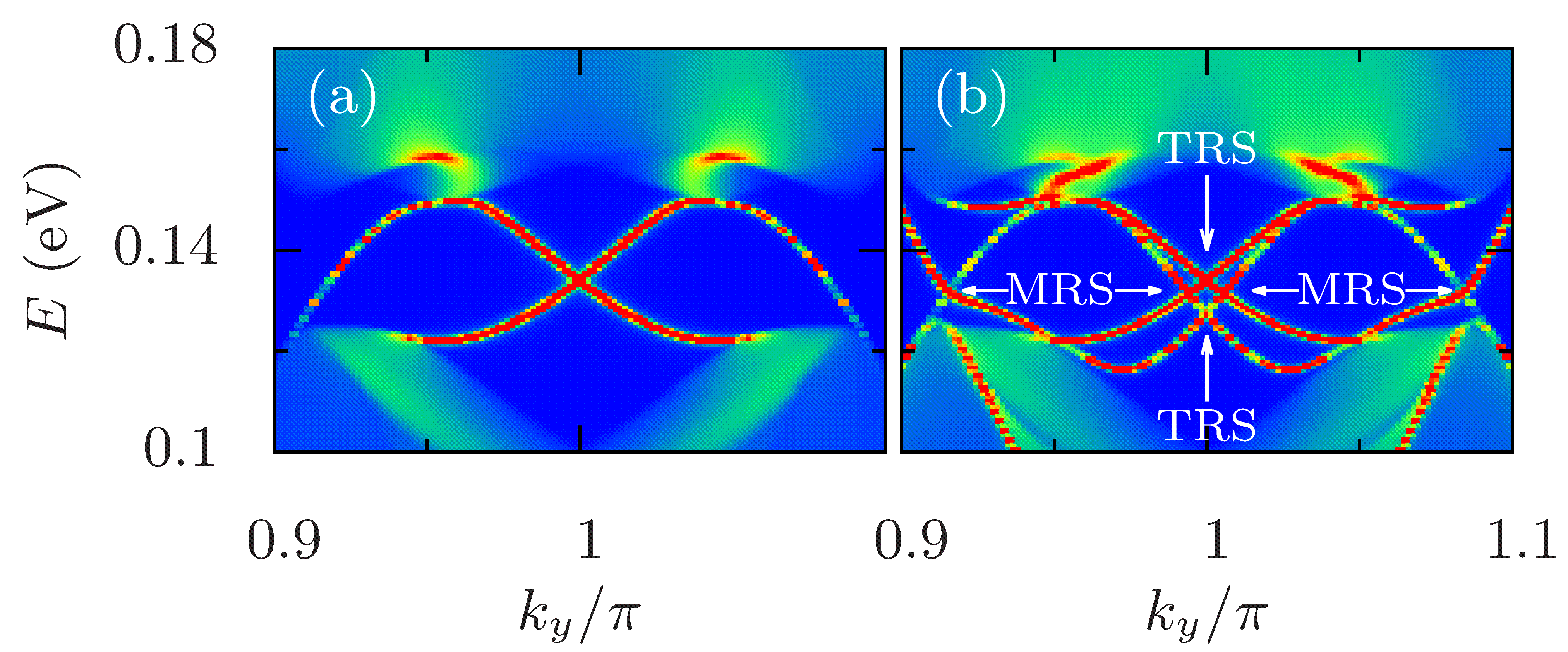}
\caption{
One-dimensional gapless edge states along the (100) edge in SnTe film with $N_z=20$ for (a) the mirror-even and (b) both sectors.
The degeneracies denoted by the arrows are protected by mirror-reflection (MRS) and time-reversal (TRS) symmetries, respectively.}
\label{nz20}
\end{figure}
Figure \ref{nz20} shows the edge charge spectral function $\rho_{+, \mathrm c}(k_y, E)$ for the mirror-even sector in the even numbers of layers of SnTe film.
There are gapless mirror-helical edge states at $k_y \sim \pi$ for $|\zeta_{\rm M}|=2$.
One can see degeneracy of the edge states at the zone boundary $k_y=\pi$ in Fig. \ref{nz20}(a), which is protected by the time-reversal symmetry defined in Appendix \ref{trs}.
Figure \ref{nz20}(b) shows the total edge-charge-spectral function $\rho_{+, \mathrm c}(k_y, E) + \rho_{-, \mathrm c}(k_y, E)$.
Degeneracies of the edge states at $k_y \ne \pi$ are protected by the (001)--reflection symmetry.
Both time-reversal and (001)--reflection symmetries are necessary for gapless edge states in the case of even numbers of layers.

\begin{figure}
\centering
\includegraphics{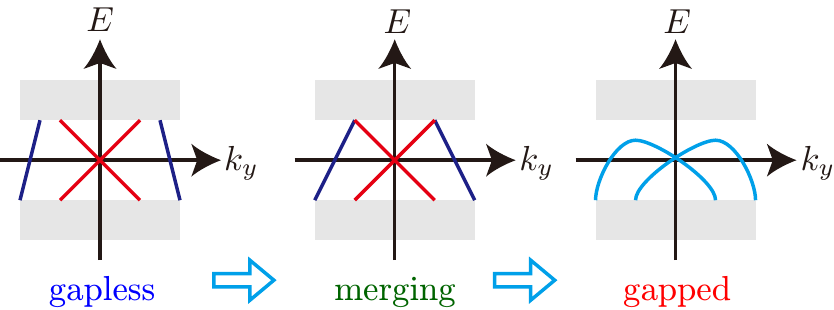}
\caption{
Evolution from gapless to gapped edge states induced by perturbation. 
The energy dispersions correspond to that in Fig. \ref{nz20}(a).}
\label{merge}
\end{figure}

The gapless edge states is, however, not robust: it is removed without bulk gap closing. 
As illustrated in Fig. \ref{merge}, two branches can be merged and become gapped continuously without any gap closing.
Although the system has the nontrivial topological invariant $\zeta_{\rm M}$,  
it has no bulk-edge correspondence.
Equation (\ref{zeta}) is a topological number protected by a complex symmetry involving spatial inversion symmetry (see Appendix \ref{symwn}), which is preserved in the bulk but not in the edge.
Namely, a nontrivial value of $\zeta_{\rm M}$ does not lead to the existence of robust gapless edge states.

The second minima of energy gap at $N_z = 40$ for the SnTe film [Fig. \ref{pd_odd}(a)] originates from the change of the topological number from $|\zeta_{\rm M}|=2$ to $\zeta_{\rm M} = 0$.
It follows that the local structure of energy dispersion for the edge states changes near the band gap, via bulk gap closing as explained below.
\begin{figure}
\centering
\includegraphics{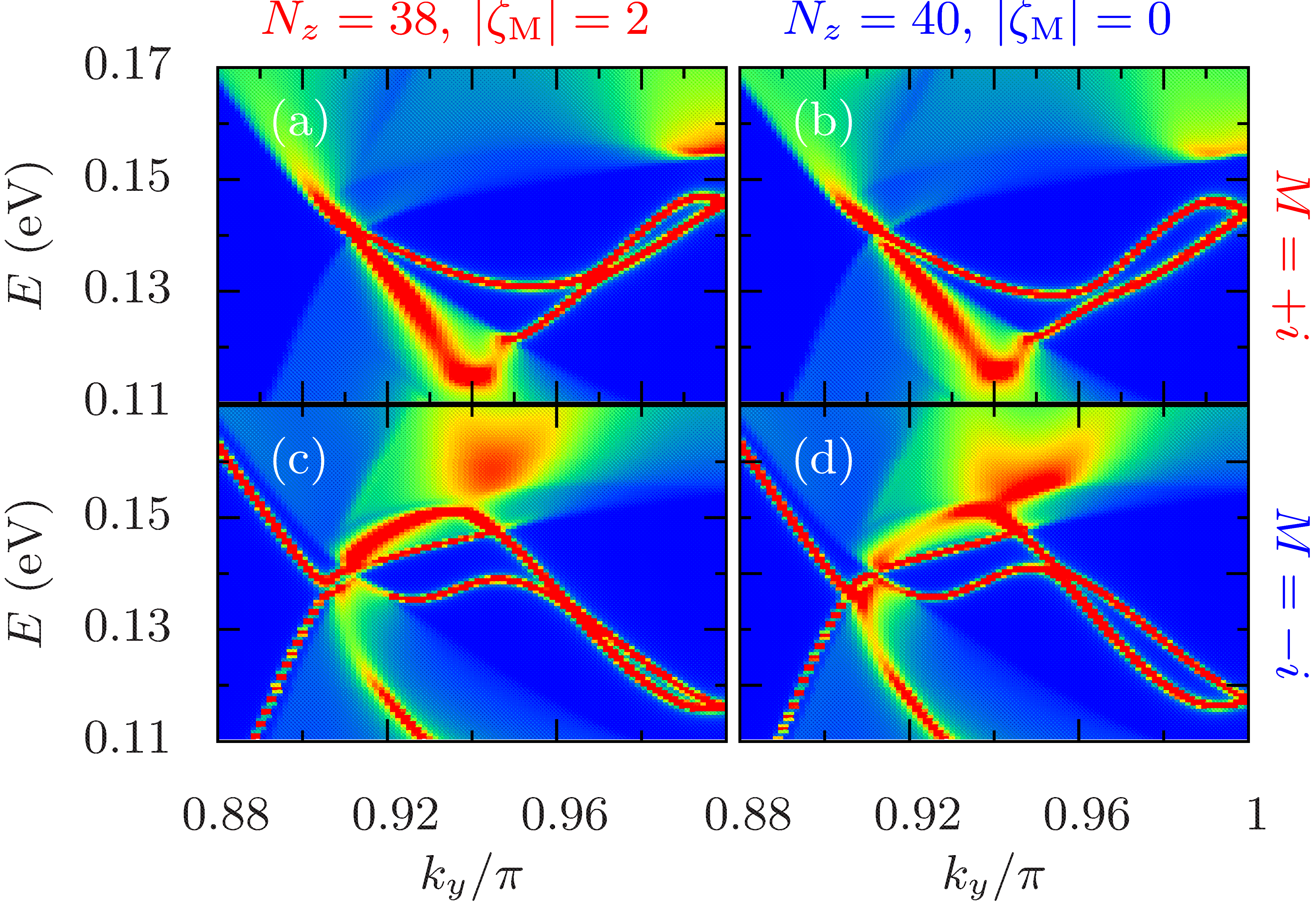}
\caption{Edge spectral function for $N_z=38$ [(a) and (c)] and $N_z=40$ [(b) and (d)]. The mirror eigenvalue is given by $M=+i$ for (a) and (b) and $M=-i$ for (c) and (d).}
\label{edge38-40}
\end{figure}
\begin{figure}
\centering
\includegraphics{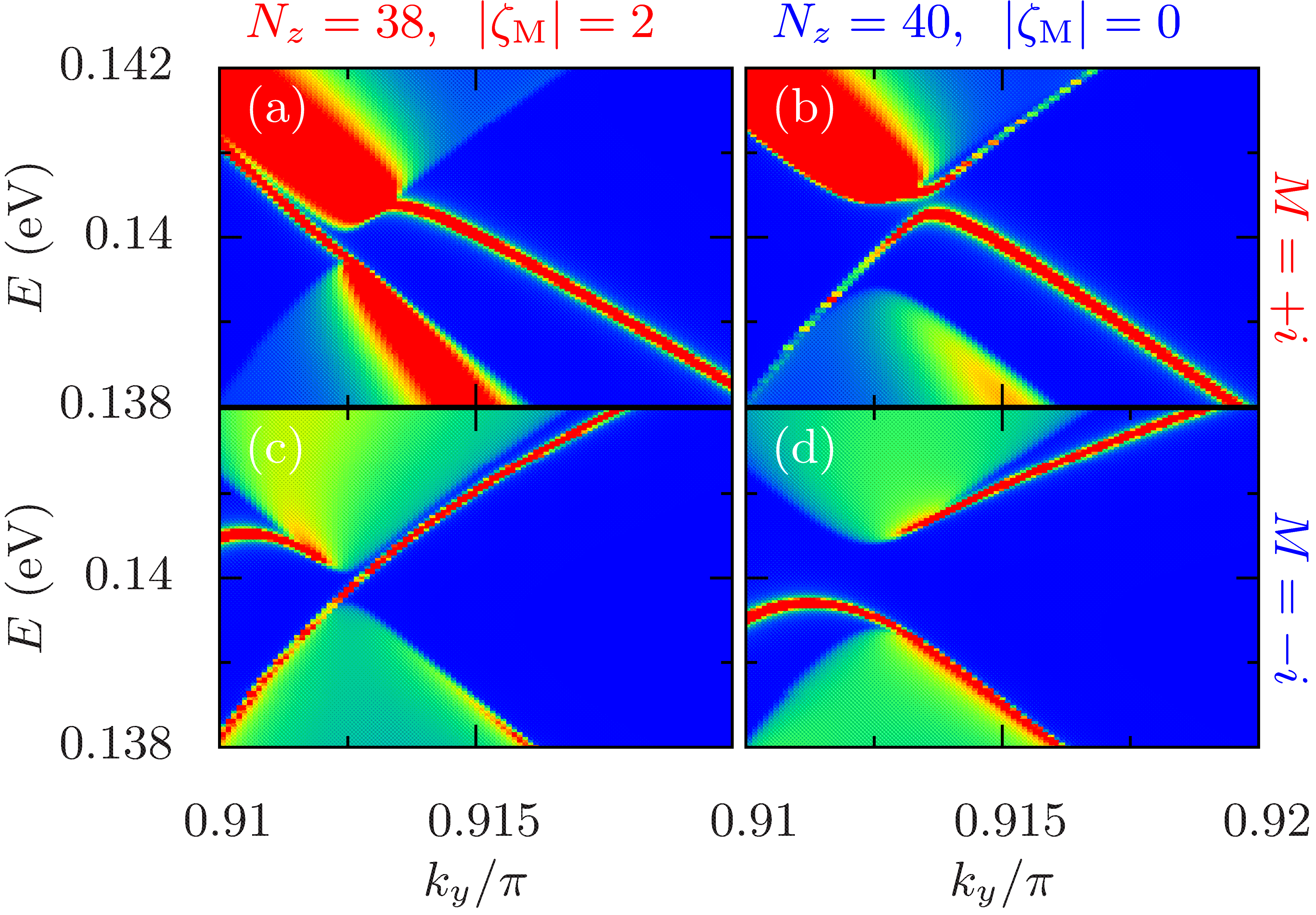}
\caption{Edge spectral function in the vicinity of the band gap. 
(a)--(d) correspond to those in Fig. \ref{edge38-40}, respectively.}
\label{edge38-40_2}
\end{figure}
Figures \ref{edge38-40} and \ref{edge38-40_2} show the edge charge spectral function for $N_z=38$ and $N_z=40$, in the vicinity of which the topological number changes from $|\zeta_{\rm M}|=2$ to $\zeta_{\rm M}=0$ and the band gap closes.
The whole structures of energy dispersion for $N_z=38$ [Figs. \ref{edge38-40}(a) and \ref{edge38-40}(c)] and $N_z=40$ [Figs. \ref{edge38-40}(b) and \ref{edge38-40}(d)] are nearly the same. 
Near the band gap ($k_y \sim 0.912 \pi$), on the other hand, gapless edge states for $N_z=38$ [Figs. \ref{edge38-40_2}(a) and \ref{edge38-40_2}(c)] become gapped for $N_z=40$ [Figs. \ref{edge38-40_2}(b) and \ref{edge38-40_2}(d)], accompanying the change of topological number $\zeta_{\rm M}$.
The change of $\zeta_{\rm M}$ can describe bulk gap closing and generation of gapless edge states, although the bulk-edge correspondence does not exactly hold.

\subsection{Discussion}

	In the thin film, the mirror Chern insulator of $|N_{\rm M}|=2$ with robust gapless mirror-chiral edge states and  $|\zeta_{\rm M}|=2$ insulator with gapless/gapful mirror-helical edge states are realized for the odd and even numbers of layers, respectively.
Experimental evidence of these edge states is quantization of two-terminal charge conductance $G_{xx}$ of the order of $G_{xx} \sim e^2/h$, e.g., $G_{xx} = 4e^2/h$ for $N_z=7$ [Fig. \ref{spin}(a)] and $G_{xx} = 8e^2/h$ for $N_z=20$ (Fig. \ref{nz20}) in the ballistic limit.

What is a quantity directly related to $N_{\rm M}$ or $\zeta_{\rm M}$?
One of the candidates is mirror Hall conductance, which is defined by
$G_{\mathrm M, xy} = G_{+, xy} - G_{-,xy}$, where $G_{\pm, xy}$ denotes charge Hall conductance in each mirror sector.
In the odd numbers of layers, the gapless mirror-chiral edge states result in the quantized mirror Hall conductance $G_{\mathrm M, xy} = 2 N_{\rm M} e^2/h$.
On the contrary, the mirror Chern number, i.e., the mirror Hall conductance vanishes ($N_{\rm M}=0$) in the even numbers of layers since the system has time-reversal symmetry in each mirror sector and the corresponding edge states are mirror-helical.
If one measure the mirror Hall conductance, the $|N_{\rm M}|=2$ and $|\zeta_{\rm M}|=2$ insulators can be distinguished from each other.

Note that it is difficult to detect the \textit{mirror} Hall conductance experimentally since \textit{mirror} is not observable.
In single band systems, the (001)--reflection operator $M$ can be identified to spin as $M = -i s_z$, i.e., mirror Hall conductance is equivalent to spin Hall conductance.
In multi-orbital systems, on the other hand, the (001)--reflection operator depends on orbitals.
Spins of each orbital can be partially canceled in each mirror sector.
In fact, we found that the gapless edge states in each mirror sector for the $|N_{\rm M}|=2$ phase has partial (about 10\%) spin polarization.
This leads to
a (not quantized) finite value of $z$-component spin Hall conductance, which is an evidence of the $|N_{\rm M}|= 2$ phase.
Besides, to find phenomena directly related to the mirror Chern number is an important remaining issue.

It is also a challenging study to characterize $|\zeta_{\rm M}| \ne 0$ insulators by physical quantities since it has no robust gapless edge state.
The relevant system is inversion symmetric insulator in odd spatial dimensions, in which the topological number does not guarantee the existence of gapless edge states but mid-gap entanglement modes\cite{turner10, hughes11} and the magneto-electric crossed response.\cite{hughes11}
In this case, the topological number is equivalent to the magneto-electric polarizability.\cite{qi08} 
Unfortunately, such a nontrivial response has not been found in even spatial dimensions.
To make matters more challenging, the topological number $\zeta_{\rm M}$ is protected by a  combination of mirror-reflection, time-reversal, translation in the reciprocal space, and inversion symmetries.
Such a crystalline topological number in even spatial dimensions also should be made a correlation to physical phenomena.


All calculations in the paper have been based on a tight-binding model which does not include the surface and edge potential effects in a self-consistent manner.
In particular, surface reconstruction could be important for  tiny energy gaps and small energy ranges over which the edge states exist. 
On the other hand, edge reconstruction is not so much important as surface reconstruction for qualitative study since the existence of edge state is guaranteed by the topological invariant in the two-dimensional bulk system.
For quantitative study in the small energy scales, surface-- and edge--reconstruction effects should be evaluated.

For transport measurement and device application, 
it is necessary to study effects of surface roughness of the film.
We have shown the non-monotonically thickness-dependent energy gap and two-dimensional topological number.
The even numbers of layers has the topological number different from that of the odd numbers of layers.
These suggest that surface roughness can crucially affect the topological phase and resulting transport phenomena.
This issue should be discussed elsewhere.

Furthermore, recently, nanowires of SnTe have been synthesized.\cite{saghir14, safdar14}
Finite-size effect of one-dimensional SnTe wires, which is natural extension of the present work, will be discussed in future work.

\section{Summary}
\label{summary}

In summary, we have elucidated electronic states and topological invariants in a TCI Pb$_{x}$Sn$_{1-x}$Te film.
Energy gap of the film shows damped oscillation as a function of the thickness.
We have also clarified that the oscillation of energy gap originates from the change of topological number from trivial to $\mathbb Z$ invariants.
This structure is richer than that of $\mathbb Z_2$ TIs such as Bi$_2$Se$_3$, in which only $\mathbb Z_2$ phase can be realized in the film.
In contrast, the topological numbers of the TCI film can be tuned to the (signed) mirror Chern number $N_{\rm M} = \pm 2$, and also to $|\zeta_{\rm M}| = 2$, by tuning the number of layers of film.
The obtained data may provide essential support for understanding electronic states and transport properties of the film, which  depend non-monotonically on the number of layers.

\begin{acknowledgments}
 The authors are grateful to Y. Ando, S. Onari,  T. Hashimoto, K. Yada, M. Ezawa, and S. Kashiwaya for fruitful discussions.
This work is supported by the ``Topological Quantum Phenomena"
(no. 22103005) Grant-in Aid for Scientific Research on Innovative
Areas from the Ministry of Education, Culture, Sports, Science and
Technology (MEXT) of Japan. 
M.S. is supported by Grant-in-Aid for
Scientific Research B (no. 25287085) from Japan Society for the
Promotion of Science (JSPS).
\end{acknowledgments}

\appendix

 \section{Model}

\subsection{(001)--film based on the 36 $\times$ 36 model}
\label{model}

Based on the $36 \times 36$ model,\cite{lent86}
we construct a model for (001)-film of IV--VI semiconductors as
\begin{align}
 H = \sum_{n_z=1}^{N_z} \bm c^\dag_{n_z} \epsilon \bm c_{n_z} + \sum_{n_z=1}^{N_z-1} \left(\bm c^\dag_{n_z} t_z \bm c_{n_z+1} + \mathrm{H.c.} \right).
\end{align}
$36 \times 36$ matrices $\epsilon$ and $t_z$ are obtained from the bulk Hamiltonian with the substitutions of $2 \cos k_z/2 \to \bm c^\dag_n \bm c_{n+1} + \bm c_{n+1}^\dag \bm c_{n}$ and $2i \sin k_z/2 \to \bm c^\dag_{n} \bm c_{n+1} - \bm c^\dag_{n+1} \bm c_{n}$.
The explicit forms of $\epsilon$ and $t_z$
are given by
\begin{align}
 \epsilon = \begin{pmatrix}
   \epsilon_{\rm ss} & \epsilon_{\rm pcs}^\dag & \epsilon_{\rm pas}^\dag & 0 & 0
   \\
   \epsilon_{\rm pcs} & \epsilon_{\rm pcpc} & \epsilon_{\rm papc}^\dag & 0 & \epsilon_{\rm dapc}^\dag
   \\
   \epsilon_{\rm pas} & \epsilon_{\rm papc} & \epsilon_{\rm papa} & \epsilon_{\rm dcpa}^\dag & 0
   \\
   0 & 0 & \epsilon_{\rm dcpa} & \epsilon_{\rm dcdc} & \epsilon_{\rm dadc}^\dag
   \\
   0 & \epsilon_{\rm dapc} & 0 & \epsilon_{\rm dadc} & \epsilon_{\rm dada}
 \end{pmatrix},
\end{align}
\begin{align}
 t_z = \begin{pmatrix}
   t_{\rm ss} & -t_{\rm pcs}^{\rm T} & -t_{\rm pas}^{\rm T} & 0 & 0
   \\
   t_{\rm pcs} & 0 & t_{\rm papc} & 0 & -t_{\rm dapc}^{\rm T}
   \\
   t_{\rm pas} & t_{\rm papc} & 0 & -t_{\rm dcpa}^{\rm T} & 0
   \\
   0 & 0 & t_{\rm dcpa} & 0 & t_{\rm dadc}
   \\
   0 & t_{\rm dapc} & 0 & t_{\rm dadc} & 0
 \end{pmatrix},
\end{align}
with
\begin{align}
 \epsilon_{\rm ss}
 =
 \begin{pmatrix}
  E_{\rm sc} & 0 & c_0 V_{\rm ss} & 0 
  \\
  0 & E_{\rm sc} & 0 & c_0 V_{\rm ss}
  \\
  c_0 V_{\rm ss} & 0 & E_{\rm sa} & 0
  \\
  0 & c_0 V_{\rm ss} & 0 & E_{\rm sa}
 \end{pmatrix},
 \end{align}
 \begin{align}
 \epsilon_{\rm pcs} =
 \begin{pmatrix}
   0 & 0 & -2 s_x V_{\rm ps} & 0
   \\
   0 & 0 & -2 s_y V_{\rm ps} & 0
   \\
   0 & 0 & 0 & 0
   \\
   0 & 0 & 0 & -2 s_x V_{\rm ps} 
   \\
   0 & 0 & 0 & -2 s_y V_{\rm ps} 
   \\
   0 & 0 & 0 & 0 
 \end{pmatrix},
 \end{align}
 \begin{align}
 \epsilon_{\rm pas} =
 \begin{pmatrix}
   -2 s_x V_{\rm sp} & 0 & 0 & 0 
   \\
   -2 s_y V_{\rm sp} & 0 & 0 & 0
   \\
   0 & 0 & 0 & 0
   \\
   0 & -2 s_x V_{\rm sp}  & 0 & 0
   \\
   0 & -2 s_y V_{\rm sp}  & 0 & 0
   \\
   0 & 0 & 0 & 0 
 \end{pmatrix},
 \end{align}
 \begin{align}
 \epsilon_{\rm pcpc} =
 \begin{pmatrix}
   E_{\rm pc} & -i \displaystyle\frac{\lambda_{\rm c}}{2} & 0 & 0 & 0 & \displaystyle\frac{\lambda_{\rm c}}{2}
   \\
   i \displaystyle\frac{\lambda_{\rm c}}{2} & E_{\rm pc} & 0 & 0 & 0 & -i\displaystyle\frac{\lambda_{\rm c}}{2}
   \\
   0 & 0 & E_{\rm pc} & - \displaystyle\frac{\lambda_{\rm c}}{2} & i \displaystyle\frac{\lambda_{\rm c}}{2} & 0
   \\
   0 & 0 & -\displaystyle\frac{\lambda_{\rm c}}{2} & E_{\rm pc} & i \displaystyle\frac{\lambda_{\rm c}}{2} & 0
   \\
   0 & 0 & -i \displaystyle\frac{\lambda_{\rm c}}{2} & -i \displaystyle\frac{\lambda_{\rm c}}{2} & E_{\rm pc} & 0
   \\
   \displaystyle\frac{\lambda_{\rm c}}{2} & i \displaystyle\frac{\lambda_{\rm c}}{2} & 0 & 0 & 0 & E_{\rm pc}
 \end{pmatrix},
 \end{align}
 \begin{align}
 \epsilon_{\rm papa} &= \epsilon_{\rm pcpc}|_{E_{\rm pc} \to E_{\rm pa}, \lambda_{\rm c} \to \lambda_{\rm a}},
 \\
 \epsilon_{\rm papc} &= \mathrm{diag}
 \left(
  W_x, W_y, c_0 V_{\mathrm {pp} \pi}, W_x, W_y, c_0 V_{\mathrm {pp} \pi}
 \right),
 \\
 \epsilon_{\rm dapc} 
 &= \mathrm{diag}
 \left(
  \epsilon_5, \epsilon_5
 \right),
\end{align}
\begin{align}
 \epsilon_5 =
 \begin{pmatrix}
   - \sqrt 3 s_x V_{\rm pd} & \sqrt 3 s_y V_{\rm pd} & 0
   \\
   s_x V_{\rm pd} & s_y V_{\rm pd} & 0
   \\
   -2 s_y V_{\mathrm{pd}\pi} & -2 s_x V_{\mathrm{pd}\pi} & 0
   \\
   0 & 0 & -2 s_y V_{\mathrm{pd}\pi}
   \\
   0 & 0 & -2 s_x V_{\mathrm{pd}\pi}
 \end{pmatrix},
 \end{align}
 \begin{align}
 \epsilon_{\rm dcpa} = \epsilon_{\rm dapc}|_{V_{\rm pd} \to V_{\rm dp}, V_{\mathrm{pd}\pi} \to V_{\mathrm{dp}\pi}}.
\end{align}
$10 \times 10$ matrix $\epsilon_{\rm dadc}$ has ten nonzero diagonal and four nonzero off-diagonal elements given by
\begin{align}
(\epsilon_{\rm dadc})_{1,1} 
= (\epsilon_{\rm dadc})_{6,6} =
 (c_x+c_y) 
\frac{3V_{\rm dd} + V_{\mathrm{dd}\delta}}{2},
\end{align}
\begin{align}
(\epsilon_{\rm dadc})_{2,2} = (\epsilon_{\rm dadc})_{7,7} =
(c_x+c_y) \frac{3V_{\mathrm{dd} \delta} + V_{\mathrm{dd}}}{2},
\end{align}
\begin{align}
(\epsilon_{\rm dadc})_{3,3} = (\epsilon_{\rm dadc})_{8,8} =
c_0 V_{\mathrm{dd}\pi},
\end{align}
\begin{align}
(\epsilon_{\rm dadc})_{4,4} = (\epsilon_{\rm dadc})_{9,9} =
2 c_y V_{\mathrm{dd}\pi} + 2 c_x V_{\mathrm{dd}\delta},
\end{align}
\begin{align}
(\epsilon_{\rm dadc})_{5,5} = (\epsilon_{\rm dadc})_{10,10} =
2 c_x V_{\mathrm{dd}\pi} + 2 c_y V_{\mathrm{dd}\delta},
\end{align}
\begin{align}
(\epsilon_{\rm dadc})_{1,2} &= (\epsilon_{\rm dadc})_{2,1} 
= (\epsilon_{\rm dadc})_{6,7} = (\epsilon_{\rm dadc})_{7,6}
\nonumber \\
&=
\frac{\sqrt 3}{2} (c_y-c_x) (V_{\rm dd} - V_{\mathrm{dd} \delta}),
\end{align}
and
\begin{align}
 c_0 = 2 \cos \frac{k_x}{2} + 2 \cos \frac{k_y}{2},
 \end{align}
\begin{align}
 c_i = \cos \frac{k_i}{2}, \ s_i = i \sin \frac{k_i}{2},
\end{align}
\begin{align}
 W_i = 2 \cos \frac{k_i}{2} V_{\rm pp} +  2 \cos \frac{k_j}{2} V_{\mathrm{pp}\pi},
\end{align}
where $i,j = x,y$ and $i \ne j$.
$\epsilon_{\rm dcdc}$ and $\epsilon_{\rm dada}$ are $10 \times 10$ diagonal matrices whose elements are given by $E_{\rm dc}$ and $E_{\rm da}$, respectively.
Matrix $t_z$ is given by
\begin{align}
\label{tss}
 t_{\rm ss} =
 \begin{pmatrix}
   0 & 0 & V_{\rm ss} & 0
   \\
   0 & 0 & 0 & V_{\rm ss}
   \\
   V_{\rm ss} & 0 & 0 & 0
   \\
   0 & V_{\rm ss} & 0 & 0
 \end{pmatrix},
 \end{align}
\begin{align}
 t_{\rm pcs} =
 \begin{pmatrix}
   0 & 0 & 0 & 0
   \\
   0 & 0 & 0 & 0
   \\
   0 & 0 & -V_{\rm ps} & 0
   \\
   0 & 0 & 0 & 0
   \\
   0 & 0 & 0 & 0
   \\
   0 & 0 & 0 & -V_{\rm ps}
 \end{pmatrix},
 \end{align}
\begin{align}
 t_{\rm pas} =
 \begin{pmatrix}
   0 & 0 & 0 & 0
   \\
   0 & 0 & 0 & 0
   \\
   -V_{\rm sp} & 0 & 0 & 0
   \\
   0 & 0 & 0 & 0
   \\
   0 & 0 & 0 & 0
   \\
   0 & -V_{\rm sp} & 0 & 0
 \end{pmatrix},
 \end{align}
\begin{align}
 t_{\rm papc} =
 \mathrm{diag}
 \left(
   V_{\mathrm{pp}\pi}, V_{\mathrm{pp}\pi}, V_{\mathrm{pp}}, V_{\mathrm{pp}\pi}, V_{\mathrm{pp}\pi}, V_{\mathrm{pp}}
 \right),
 \end{align}
\begin{align}
 t_{\rm dapc} =
 \mathrm{diag}
 \left(
   t_5, t_5
 \right),
 \end{align}
\begin{align}
 t_5 =
 \begin{pmatrix}
   0 & 0 & 0
   \\
   0 & 0 & -V_{\rm pd}
   \\
   0 & 0 & 0
   \\
   0 & -V_{\mathrm{pd}\pi} & 0
   \\
   -V_{\mathrm{pd}\pi} & 0 & 0
 \end{pmatrix},
 \end{align}
\begin{align}
 t_{\rm dcpa} = t_{\rm dapc}|_{V_{\rm pd} \to V_{\rm dp}, V_{\mathrm{pd}\pi} \to V_{\mathrm{dp}\pi}},
 \end{align}
\begin{align}
 t_{\rm dadc} &=
 \mathrm{diag}
 \left(
   V_{\mathrm{dd}\delta}, 
   V_{\rm dd}, 
   V_{\mathrm{dd}\delta}, 
   V_{\mathrm{dd}\pi}, 
   V_{\mathrm{dd}\pi}, 
 \right. \nonumber\\ & \qquad\qquad \left.
   V_{\mathrm{dd}\delta}, 
   V_{\rm dd}, 
   V_{\mathrm{dd}\delta}, 
   V_{\mathrm{dd}\pi}, 
   V_{\mathrm{dd}\pi}
 \right).
\end{align}
Here,
the basis is taken to be 
(
$\left| s, \mathrm c, \uparrow \right\rangle$, 
$\left| s, \mathrm c, \downarrow \right\rangle$, 
$\left| s, \mathrm a, \uparrow \right\rangle$, 
$\left| s, \mathrm a, \downarrow \right\rangle$, 
$\left| p_x, \mathrm c, \uparrow \right\rangle$,
$\left| p_y, \mathrm c, \uparrow \right\rangle$,
$\left| p_z, \mathrm c, \uparrow \right\rangle$,
$\left| p_x, \mathrm c, \downarrow \right\rangle$,
$\left| p_y, \mathrm c, \downarrow \right\rangle$,
$\left| p_z, \mathrm c, \downarrow \right\rangle$,
$\left| p_x, \mathrm a, \uparrow \right\rangle$,
$\left| p_y, \mathrm a, \uparrow \right\rangle$,
$\left| p_z, \mathrm a, \uparrow \right\rangle$,
$\left| p_x, \mathrm a, \downarrow \right\rangle$,
$\left| p_y, \mathrm a, \downarrow \right\rangle$,
$\left| p_z, \mathrm a, \downarrow \right\rangle$,
$\left| d_{x^2-y^2}, \mathrm c, \uparrow \right\rangle$,
$\left| d_{3z^2-r^2}, \mathrm c, \uparrow \right\rangle$,
$\left| d_{xy}, \mathrm c, \uparrow \right\rangle$,
$\left| d_{yz}, \mathrm c, \uparrow \right\rangle$,
$\left| d_{zx}, \mathrm c, \uparrow \right\rangle$,
$\left| d_{x^2-y^2}, \mathrm c, \downarrow \right\rangle$,
$\left| d_{3z^2-r^2}, \mathrm c, \downarrow \right\rangle$,
$\left| d_{xy}, \mathrm c, \downarrow \right\rangle$,
$\left| d_{yz}, \mathrm c, \downarrow \right\rangle$,
$\left| d_{zx}, \mathrm c, \downarrow \right\rangle$,
$\left| d_{x^2-y^2}, \mathrm a, \uparrow \right\rangle$,
$\left| d_{3z^2-r^2}, \mathrm a, \uparrow \right\rangle$,
$\left| d_{xy}, \mathrm a, \uparrow \right\rangle$,
$\left| d_{yz}, \mathrm a, \uparrow \right\rangle$,
$\left| d_{zx}, \mathrm a, \uparrow \right\rangle$,
$\left| d_{x^2-y^2}, \mathrm a, \downarrow \right\rangle$,
$\left| d_{3z^2-r^2}, \mathrm a, \downarrow \right\rangle$,
$\left| d_{xy}, \mathrm a, \downarrow \right\rangle$,
$\left| d_{yz}, \mathrm a, \downarrow \right\rangle$,
$\left| d_{zx}, \mathrm a, \downarrow \right\rangle$
).
In the actual calculations, 
the parameters are taken from Ref. \onlinecite{lent86}.

\subsection{Model for (001)--film with (100)--edge}
\label{edgemodel}

A model Hamiltonian for the film with (100)--edge can be obtained in a similar manner to the previous case: substitution of $2\cos k_x/2 \to c^\dag_{n_x, n_z}(k_y) c_{n_x+1, n_z}(k_y) + c^\dag_{n_x+1, n_z}(k_y) c_{n_x, n_z}(k_y)$ and $2 i \sin k_x/2 \to c^\dag_{n_x, n_z}(k_y) c_{n_x+1, n_z}(k_y) - c^\dag_{n_x+1, n_z}(k_y) c_{n_x, n_z}(k_y)$.
The resulting Hamiltonian has the form as
\begin{align}
 \tilde H &= \sum_{n_z=1}^{N_z} \sum_{n_x=1}^{N_x}
 c^\dag_{n_z, n_x} \tilde \epsilon c_{n_z, n_x}
 \nonumber \\ &
 +
 \sum_{n_z=1}^{N_z-1} \sum_{n_x=1}^{N_x} 
 c^\dag_{n_z, n_x} t_z c_{n_z+1, n_x}
 + \mathrm{H.c.}
 \nonumber \\ &
 +
 \sum_{n_z=1}^{N_z} \sum_{n_x=1}^{N_x-1} 
 c^\dag_{n_z, n_x} \tilde t_x c_{n_z, n_x+1}
 + \mathrm{H.c.}.
\end{align}
The on-site energy $\tilde \epsilon$ is given by
\begin{align}
 \tilde \epsilon &= \begin{pmatrix}
   \tilde \epsilon_{\rm ss} & \tilde \epsilon_{\rm pcs}^\dag & \tilde \epsilon_{\rm pas}^\dag & 0 & 0
   \\
   \tilde\epsilon_{\rm pcs} & \epsilon_{\rm pcpc} & \tilde\epsilon_{\rm papc}^\dag & 0 & \tilde\epsilon_{\rm dapc}^\dag
   \\
   \tilde\epsilon_{\rm pas} & \tilde\epsilon_{\rm papc} & \epsilon_{\rm papa} & \tilde\epsilon_{\rm dcpa}^\dag & 0
   \\
   0 & 0 & \tilde\epsilon_{\rm dcpa} & \tilde\epsilon_{\rm dcdc} & \tilde\epsilon_{\rm dadc}^\dag
   \\
   0 & \tilde\epsilon_{\rm dapc} & 0 & \tilde\epsilon_{\rm dadc} & \tilde\epsilon_{\rm dada}
 \end{pmatrix},
\end{align}
with
\begin{align}
 \tilde\epsilon_{\rm ss}
 =
 \begin{pmatrix}
  E_{\rm sc} & 0 & 2 c_y V_{\rm ss} & 0 
  \\
  0 & E_{\rm sc} & 0 & 2 c_y V_{\rm ss}
  \\
  2 c_y V_{\rm ss} & 0 & E_{\rm sa} & 0
  \\
  0 & 2 c_y V_{\rm ss} & 0 & E_{\rm sa}
 \end{pmatrix},
\end{align}
\begin{align}
 \tilde\epsilon_{\rm pcs} =
 \begin{pmatrix}
   0 & 0 & 0 & 0
   \\
   0 & 0 & -2 s_y V_{\rm ps} & 0
   \\
   0 & 0 & 0 & 0
   \\
   0 & 0 & 0 & 0 
   \\
   0 & 0 & 0 & -2 s_y V_{\rm ps} 
   \\
   0 & 0 & 0 & 0 
 \end{pmatrix},
 \end{align}
 \begin{align}
 \tilde\epsilon_{\rm pas} =
 \begin{pmatrix}
   0 & 0 & 0 & 0 
   \\
   -2 s_y V_{\rm sp} & 0 & 0 & 0
   \\
   0 & 0 & 0 & 0
   \\
   0 & 0  & 0 & 0
   \\
   0 & -2 s_y V_{\rm sp}  & 0 & 0
   \\
   0 & 0 & 0 & 0 
 \end{pmatrix},
 \end{align}
 \begin{align}
   \tilde \epsilon_{\rm papc} = 
2c_y
\mathrm{diag}
 \left(
  V_{\mathrm{pp}\pi}, 
  V_{\rm pp}, 
  V_{\mathrm {pp} \pi}, 
  V_{\mathrm{pp}\pi}, 
  V_{\rm pp}, 
  V_{\mathrm {pp} \pi}
 \right),
 \end{align}
 \begin{align}
 \tilde \epsilon_{\rm dapc} 
 = \mathrm{diag}
 \left(
  \tilde \epsilon_5, \tilde \epsilon_5
 \right),
\end{align}
\begin{align}
 \tilde \epsilon_5 =
 \begin{pmatrix}
   0 & \sqrt 3 s_y V_{\rm pd} & 0
   \\
   0 & s_y V_{\rm pd} & 0
   \\
   -2 s_y V_{\mathrm{pd}\pi} & 0 & 0
   \\
   0 & 0 & -2 s_2 V_{\mathrm{pd}\pi}
   \\
   0 & 0 & 0
 \end{pmatrix},
 \end{align}
 \begin{align}
 \tilde \epsilon_{\rm dcpa} &= 
\tilde\epsilon_{\rm dapc}|_{V_{\rm pd} \to V_{\rm dp}, V_{\mathrm{pd}\pi} \to V_{\mathrm{dp}\pi}}.
\end{align}

\begin{align}
(\tilde \epsilon_{\rm dadc})_{1,1} &= (\tilde \epsilon_{\rm dadc})_{6,6} =
 c_y 
\frac{3V_{\rm dd} + V_{\mathrm{dd}\delta}}{2},
\\
(\tilde \epsilon_{\rm dadc})_{2,2} &= (\tilde \epsilon_{\rm dadc})_{7,7} =
c_y \frac{3V_{\mathrm{dd} \delta} + V_{\mathrm{dd}}}{2},
\\
(\tilde \epsilon_{\rm dadc})_{3,3} &= (\tilde \epsilon_{\rm dadc})_{8,8} =
2c_y V_{\mathrm{dd}\pi},
\\
(\tilde \epsilon_{\rm dadc})_{4,4} &= (\tilde \epsilon_{\rm dadc})_{9,9} =
2 c_y V_{\mathrm{dd}\pi},
\\
(\tilde \epsilon_{\rm dadc})_{5,5} &= (\tilde\epsilon_{\rm dadc})_{10,10} =
2 c_y V_{\mathrm{dd}\delta},
\\
(\tilde \epsilon_{\rm dadc})_{1,2} &= (\tilde \epsilon_{\rm dadc})_{2,1} 
= (\tilde \epsilon_{\rm dadc})_{6,7} = (\tilde \epsilon_{\rm dadc})_{7,6}
\nonumber \\
&=
\frac{\sqrt 3}{2} c_y (V_{\rm dd} - V_{\mathrm{dd} \delta}).
\end{align}
The hopping along (001)--direction $t_z$ is the same as that defined in the previous section.
The hopping along (100)--direction $t_x$ is given by
\begin{align}
t_x &= \begin{pmatrix}
   t_{\rm ss} & -t_{x, \rm pcs}^{\rm T} & -t_{x, \rm pas}^{\rm T} & 0 & 0
   \\
   t_{x, \rm pcs} & 0 & t_{x, \rm papc} & 0 & -t_{x, \rm dapc}^{\rm T}
   \\
   t_{x, \rm pas} & t_{x, \rm papc} & 0 & -t_{x, \rm dcpa}^{\rm T} & 0
   \\
   0 & 0 & t_{x, \rm dcpa} & 0 & t_{x, \rm dadc}
   \\
   0 & t_{x, \rm dapc} & 0 & t_{x, \rm dadc} & 0
 \end{pmatrix},
\end{align}
with $t_{\rm ss}$ defined by Eq. (\ref{tss}) and
\begin{align}
 t_{x, \rm pcs} =
 \begin{pmatrix}
   0 & 0 & -V_{\rm ps} & 0
   \\
   0 & 0 & 0 & 0
   \\
   0 & 0 & 0 & 0
   \\
   0 & 0 & 0 & -V_{\rm ps}
   \\
   0 & 0 & 0 & 0
   \\
   0 & 0 & 0 & 0
 \end{pmatrix},
 \end{align}
 \begin{align}
 t_{x, \rm pas} =
 \begin{pmatrix}
   -V_{\rm sp} & 0 & 0 & 0
   \\
   0 & 0 & 0 & 0
   \\
   0 & 0 & 0 & 0
   \\
   0 & -V_{\rm sp} & 0 & 0
   \\
   0 & 0 & 0 & 0
   \\
   0 & 0 & 0 & 0
 \end{pmatrix},
 \end{align}
 \begin{align}
 t_{x, \rm papc} =
 \mathrm{diag}
 \left(
   V_{\mathrm{pp}}, V_{\mathrm{pp}\pi}, V_{\mathrm{pp} \pi}, 
   V_{\mathrm{pp}}, V_{\mathrm{pp}\pi}, V_{\mathrm{pp} \pi}
 \right),
 \end{align}
 \begin{align}
 t_{x, \rm dapc} =
 \mathrm{diag}
 \left(
   t_{x,5}, t_{x,5}
 \right),
 \end{align}
 \begin{align}
 t_{x, 5} =
 \begin{pmatrix}
   - {\sqrt 3} V_{\rm pd}/2 & 0 & 0
   \\
   V_{\rm pd}/2 & 0 & 0
   \\
   0 & - V_{\mathrm{pd}\pi} & 0
   \\
   0 & 0 & 0
   \\
   0 & 0 & -V_{\mathrm{pd}\pi}
 \end{pmatrix},
 \end{align}
 \begin{align}
  t_{x, \rm dcpa} = t_{x, \rm dapc}|_{V_{\rm pd} \to V_{\rm dp}, V_{\mathrm{pd}\pi} \to V_{\mathrm{dp}\pi}},
 \end{align}
 \begin{align}
 (t_{x, \rm dadc})_{1,1} &=
 (t_{x, \rm dadc})_{6,6} 
 =
 \frac{3 V_{\rm dd} + V_{\mathrm{dd}\delta}}{4},
 \\
 (t_{x, \rm dadc})_{2,2} &=
 (t_{x, \rm dadc})_{7,7} 
 =
 \frac{3 V_{\mathrm{dd} \delta} + V_{\mathrm{dd}}}{4},
 \\
 (t_{x, \rm dadc})_{3,3} &=
 (t_{x, \rm dadc})_{8,8} 
 =
 V_{\mathrm{dd}\pi},
 \\
 (t_{x, \rm dadc})_{4,4} &=
 (t_{x, \rm dadc})_{9,9} 
 =
 V_{\mathrm{dd}\delta},
 \\
 (t_{x, \rm dadc})_{5,5} &=
 (t_{x, \rm dadc})_{10,10} 
 =
 V_{\mathrm{dd}\pi},
 \\
 (t_{x, \rm dadc})_{1,2} &= (t_{x, \rm dadc})_{2,1} = (t_{x, \rm dadc})_{6,7} = (t_{x, \rm dadc})_{7,6} 
\nonumber \\
&= -\frac{\sqrt 3}{4} (V_{\rm dd} - V_{\rm dd \delta}).
\end{align}

\subsection{(111)--film of Bi$_2$Se$_3$}
\label{bi2se3}

Here we supply a tight-binding model of (111)--film of TI Bi$_2$Se$_3$.
The Hamiltonian in low-energy region consists of two orbitals and spins as\cite{ebihara12}
\begin{align}
 H_{\rm TI} = \sum_{n=1}^{N_z} 
c^\dag_n \epsilon_{\rm TI} c_n
+ 
\left(
\sum_{n=1}^{N_z-1}
c^\dag_n t_{z \rm TI} c_{n+1} + \mathrm{h.c.}
\right),
\end{align}
with
\begin{align}
 \epsilon_{\rm TI}
 = \begin{pmatrix}
   C + M & 0 & B \sin k_z c & A_-
   \\
   0 & C+M & A_+ & -B \sin k_z c
   \\
   B \sin k_z c & A_- & C-M & 0
   \\
   A_+ & -B \sin k_z c & 0 & C-M
 \end{pmatrix},
\end{align}
\begin{align}
 t_{z \rm TI}
 = \begin{pmatrix}
   -M_1-C_1 & 0 & i B/2 & 0
   \\
   0 & -M_1 -C_1 & 0 & -iB/2
   \\
   iB/2 & 0 & M_1-C_1 & 0
   \\
   0 & -iB/2 & 0 & M_1-C_1
 \end{pmatrix},
\end{align}
\begin{align}
 C &= 2 C_1 + C_2 (2- \cos k_x a - \cos k_y a),
\\ 
 M &= M_0 + 2M_1+M_2(2- \cos k_x a - \cos k_y a),
 \\
 A_\pm &= A (\sin k_x a \pm i \sin k_y a),
\end{align}
where 
$N_z$ is the number of the quintuple layers of film, and
the basis is taken as $(| +, \uparrow \rangle, |+, \downarrow \rangle, |-, \uparrow \rangle, |-, \downarrow \rangle )$, in which $\pm$ and $\uparrow$/$\downarrow$ denote the parity eigenvalue of the orbital and spin respectively.
The parameters for Bi$_2$Se$_3$ are evaluated as follows,\cite{ebihara12} $M_0 = -0.28$eV, $M_1 = 0.216$eV, $M_2 = 2.60$eV, $A = 0.80$eV, $B=0.32$eV, $C_1=0.024$eV, $C_2=1.77$eV, $a=4.14$\AA, and $c=9.55$\AA.
In the model the crystal structure is approximated to be cubic while the actual structure is rhombohedral, since the difference does not alter the long-wavelength physics.

\section{Symmetry}
\label{symmetry}

Symmetries of the film is different from those of the three-dimensional bulk system.
Moreover, the even and odd numbers of layers have different symmetries.
The result is summarized in TABLE \ref{tab1}.
\begin{table}
\centering
\begin{tabular}{c|cc}
\hline\hline
 \# of layer & even & odd
 \\
 class & AII--R$_+$ & AII--R$_-$
 \\
 & $[\hat \Theta_{\bar{\rm X}}, \sqrt{\mathcal M^2} \mathcal M] = 0$ & $\{ \hat \Theta, \sqrt{\mathcal M^2} \mathcal M \} = 0$
 \\
  \hline
  period & $4\pi$ & $2\pi$
 \\
 Kramers & $\rm \bar X$, $\rm \bar M$ & no
 \\
inversion & yes & yes
\\
 \hline\hline
\end{tabular}
\caption{Summary of symmetries of the film.
Symmetry classes of the even and odd numbers of layers are AII--R$_+$ and AII--R$_-$ defined in Refs. \onlinecite{chiu13, morimoto13}, respectively.
Time-reversal operators $\hat\Theta$ and $\hat\Theta_{\rm \bar X}$ are defined in Appendix \ref{trs}.
$\mathcal M^2$ is proportional to the identity matrix.
Period in the momentum space, position of Kramers pair (time-reversal invariant momentum), are summarized for the even and odd numbers of layers in each mirror sector.
}
\label{tab1}
\end{table}


\subsection{(001)--reflection symmetry}
\label{001}

First, we start with (001)--reflection symmetry, which is satisfied in both the even and odd numbers of layers.
The Hamiltonian has the form
\begin{align}
 H(\bm k) &= \sum_{n=1}^{N_z} c^\dag_n(\bm k) \epsilon(\bm k) c_n(\bm k) 
\nonumber \\ & \hspace{-1em}
+ \sum_{n=1}^{N_z-1} 
\left[
c^\dag_n(\bm k) t_z c_{n+1}(\bm k)
+
c_{n+1}^\dag(\bm k) t_z^\dag c_n(\bm k)
\right],
\end{align}
where $\bm k = (k_x, k_y)$, matrices $\epsilon$ and $t_z$ are defined in Appendix \ref{model}.
$\epsilon$ and $t_z$ are transformed by (001)--reflection, as follows,
\begin{align}
 M \epsilon(\bm k) M^\dag = \epsilon(\bm k),
 \
 M t_z M^\dag = t_z^\dag.
\end{align}
(001)--reflection operator $M$ is given by $M = -i P_z s_z$, where $P_z$ is the $z \to -z$ operator acting on the orbital space and $s_z$ is the $z$--component of spin.
From the above relations, one finds the (001)--reflection symmetry $\mathcal M H(\bm k) \mathcal M^\dag = H(\bm k)$ with
\begin{align}
  \mathcal M c_{n}(\bm k) \mathcal M^\dag = \eta M c_{N_z+1-n}(\bm k),
\end{align}
where phase factor $\eta$ is defined in Eq. (\ref{def1}).
Note that the above discussion is applied also for the even numbers of layers, nevertheless the actual lattice with even number of layers does not have (001)--reflection symmetry.
This is because the system has in-plane translational symmetry. 
The Hamiltonian $H$ is expressed in the momentum $(k_x, k_y)$ space so that the translational symmetry is implemented in $H$. 
Namely, the microscopic positions of cation and anion are no longer distinguished in $H(k_x, k_y)$. 
In this sense, the even number of layers has the (001)--reflection symmetry.

With the help of the (001)--reflection symmetry,
Hamiltonian $H$ can be decomposed into two mirror sectors: 
\begin{align}
H = H_+ \oplus H_-,
\
H_{\pm} = P_{\pm} H P_{\pm},
\end{align}
 with the mirror-projection operator $P_\pm = \sum_m |m, \pm \rangle \langle m, \pm |$, 
$\mathcal M | m, \pm \rangle = \pm | m, \pm \rangle$,  and $|m, \pm \rangle$ denotes the $m$--th eigenvector of $H_\pm$.

\subsection{Periodicity}
\label{periodicity}

The projected Brillouin zone is given by $|k_y| \leq 2\pi - |k_x|$, $k_x \in [-2 \pi, 2 \pi]$.
The reciprocal lattice vectors are defined by $\bm G_1 = 2 \pi (\bm e_x + \bm e_y)$ and 
$\bm G_2 = 2 \pi (-\bm e_x + \bm e_y)$ with $\bm e_x = (1,0)$ and $\bm e_y = (0,1)$.
However, the Hamiltonian does not have the trivial periodicity: 
$H(\bm k) \ne H(\bm k + \bm G_i)$.
In order to retain the periodicity,
a gauge transformation $U$ is needed.
Periodicity of $\epsilon$ and $t_z$ gives following relations,
\begin{align}
 \epsilon(\bm k) = U \epsilon(\bm k + \bm G_i) U^\dag, 
\
t_z = - U t_z U^\dag,
\end{align}
where $U$ is the operator giving a negative sign to wave functions of the anions.
Consequently, periodicity of the film is satisfied as
\begin{align}
 H(\bm k) = \mathcal U H(\bm k + \bm G_i) \mathcal U^\dag,
\end{align}
with
\begin{align}
 \mathcal U c_{n}(\bm k) \mathcal U^\dag = (-1)^n U c_{n} (\bm k + \bm G_i).
\end{align}
The negative sign factor changes the periodicity in each mirror sector: in the case of odd (even) numbers of layers, $\mathcal U$ (anti)commutes with $\mathcal M$.
\begin{align}
 \mathcal M \mathcal U \mathcal + (-1)^{N_z} \mathcal U \mathcal M = 0.
\end{align}
Thus $2\pi$ periodicity $H_{\pm}(\bm k) = \mathcal U H_{\pm}(\bm k + \bm G_i) \mathcal U^\dag$ is satisfied only for the odd numbers of layers.
On the other hand, for the even numbers of layers, the Hamiltonian has the doubled periodicity 
\begin{align}
H_{\pm}(\bm k) = \mathcal U H_{\mp}(\bm k + \bm G_i) \mathcal U^\dag = H_{\pm}(\bm k + \bm G_1 + \bm G_2).
\label{period2}
\end{align}

\begin{figure}
\centering
\includegraphics{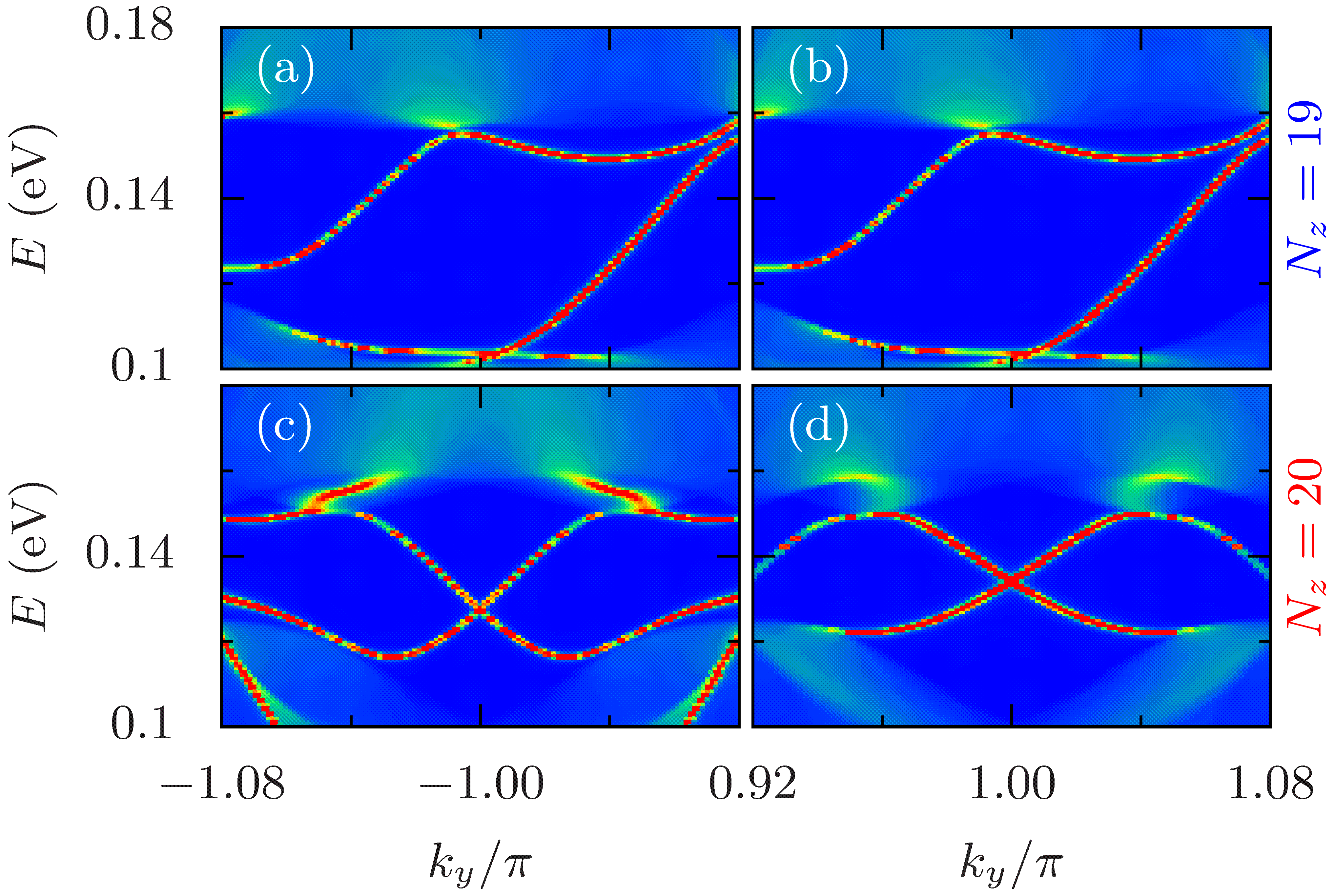}
\caption{$2\pi$ [(a) and (b)] and $4\pi$ [(c) and (d)] periodicities of energy dispersion in each mirror sector. 
Edge spectral function for $N_z=19$ [(a) and (b)] and for $N_z=20$ [(c) and (d)] of SnTe film.
(a) and (c) [(b) and (d)] show those around $k_y=-\pi$ ($k_y=\pi$).
}
\label{period}
\end{figure}
The periodicities can be directly confirmed in the (10)--edge spectral functions shown in Fig. \ref{period}.
The edge spectral function for the odd numbers of layers in $-2\pi \leq k_y \leq 0$ [Fig. \ref{period}(a)] takes the same value as that in 
$0 \leq k_y \leq 2\pi$ [Fig. \ref{period}(b)], i.e., the period is $2\pi$.
 On the other hand, for the even numbers of layers, those in $-2\pi \leq k_y \leq 0$ [Fig. \ref{period}(c)] and in $0 \leq k_y \leq 2\pi$ [Fig. \ref{period}(d)] are different, i.e., the resulting period is given by not $2\pi$ but $4\pi$.

\subsection{Time-reversal symmetry}
\label{trs}

The system has time-reversal symmetry,
\begin{align}
\label{trs1}
 H(\bm k) &= \hat \Theta H(-\bm k) \hat \Theta^{-1},
 \\
 \hat \Theta c_n(\bm k) \hat \Theta^{-1} 
&= \Theta c_n(-\bm k),
 \\
 \hat \Theta c_n^\dag(\bm k) \hat \Theta^{-1} 
&=  c_n^\dag(-\bm k) \Theta^{-1},
\end{align}
with $\Theta = -i s_y \mathcal K$.
Matrices $\epsilon(\bm k)$ and $t_z$ has the following relation,
\begin{align}
 \Theta \epsilon(\bm k) \Theta^{-1} = \epsilon(-\bm k),
 \
 \Theta t_z \Theta^{-1} = t_z.
\end{align}
The $\bar \Gamma$ and $\rm \bar M$ points are time-reversal invariant momenta (TRIM),
but the $\bar {\rm X}$ point is not.
At $\rm \bar {X}$: $(\pi, \pi) = \bm G_1/2$ and
$\rm \bar Y$:  
$(-\pi, \pi) = \bm G_2/2$ points, 
the different time-reversal symmetry is satisfied as
\begin{align}
\label{trs2}
 H(\bm k + \bm G_i/2) 
= \hat \Theta_{\rm \bar X} H(-\bm k + \bm G_i/2) \hat \Theta_{\rm \bar X}^{-1},
\end{align}
with $\hat \Theta_{\rm \bar X} = \mathcal U \hat \Theta$.
Similarly to the periodicity in the previous section, the negative sign factor changes the symmetry in each mirror sector.
$\hat \Theta_{\rm \bar X}$ (anti)commutes with $\mathcal M$ for odd (even) numbers of layers;
\begin{align}
\label{comm_MT}
 \mathcal M \hat \Theta_{\rm \bar X} + (-1)^{N_z} \hat \Theta_{\rm \bar X} \mathcal M = 0,
\end{align}
i.e., the system belongs to the class AII--U$_-^+$ in Ref. \onlinecite{shiozaki14} (or AII--R$_-$ in Refs. \onlinecite{chiu13, morimoto13}) in odd numbers of layers and to the class AII--U$_-^-$ (or AII--R$_+$) in even numbers of layers.
As a result, in the even numbers of layers, $\hat \Theta_{\rm \bar X}$ symmetry requires Kramers pairs only at the $\rm \bar X$ and $\rm \bar Y$ points within each mirror sector.
Note that $\Theta$ interchanges the mirror sectors; $\hat \Theta H_\pm(\bm k) \hat \Theta^{-1} = H_\mp(-\bm k)$, due to $[ \hat \Theta, \mathcal M ] = 0$.

\subsection{Spatial-inversion symmetry}
\label{inv}

As in the case of the time-reversal symmetry, spatial-inversion symmetry takes different forms depending on the TRIM.
At the $\rm \bar \Gamma$ point,
the inversion symmetry is expressed as follows,
\begin{align}
 H(\bm k) &= \mathcal P H(-\bm k) \mathcal P^\dag,
 \\
 \mathcal P c_n(\bm k) \mathcal P^\dag
&= P c_{N_z + 1 -n}(-\bm k),
\end{align}
where $P$ acts on the orbital space.
$\mathcal P$ commutes with $\mathcal M$ for both even and odd numbers of layers.
On the other hand,
at the $\rm \bar X$ point,
\begin{align}
 H(\bm k + \bm G_i/2) = \mathcal P_{\rm \bar X} H(-\bm k + \bm G_i/2) \mathcal P_{\rm \bar X}^\dag,
\label{PX}
\end{align}
with $\mathcal P_{\rm \bar X} = \mathcal U \mathcal P$.
$\mathcal P_{\rm \bar X}$ (anti)commutes with $\mathcal M$ for the odd (even) numbers of layers; 
\begin{align}
\label{comm_MP}
 \mathcal M \mathcal P_{\rm \bar X} + (-1)^{N_z} \mathcal P_{\rm \bar X} \mathcal M = 0.
\end{align}

Equations (\ref{comm_MT}) and (\ref{comm_MP}) lead to
\begin{align}
 \mathcal P \hat \Theta H_\pm \Theta^{-1} \mathcal P^\dag = H_\mp,
 \
  \mathcal P_{\rm \bar X} \hat \Theta_{\rm \bar X} 
H_\pm \Theta^{-1}_{\rm \bar X} \mathcal P^\dag_{\rm \bar X} = H_\mp,
\end{align}
namely, all the energy bands in each sector is not doubly degenerated, except for the $\rm \bar X$ points.

\subsection{Rotational symmetry}

The film has $C_4$ symmetry along the $z$-axis:
\begin{align}
 H(\bm k) 
= \mathcal C_4  H(\bm k') \mathcal C_4^\dag ,
 \
 \mathcal C_4 c_n(\bm k) \mathcal C_4^\dag = C_4 c_{n}(\bm k')
\end{align}
with $C_4 = e^{-i j_z \pi/2}$,
where $\bm k'$ is the rotated momentum and $j_z$ denotes the $z$-component of the total angular momentum. 
Combination of $C_4$, $\Theta$, and $U$ leads to
\begin{align}
 H\left(\bm k + \bm{\bar{M}} \right) 
= \tilde { \mathcal{C}}_4  
H\left(-\bm k' + \bm{\bar M} \right) \tilde{ \mathcal C}_4^\dag,
\end{align}
with  $\tilde {\mathcal C}_4 = \mathcal C_4 \mathcal U \hat \Theta$ and $\bm{\bar M} = (\bm G_1 + \bm G_2)/2$ being the $\rm \bar M$ point.
Furthermore, $\tilde {\mathcal C}_4$ (anti)commutes with $\mathcal M$ for odd (even) numbers of layers;
\begin{align}
 \mathcal M \tilde {\mathcal C}_4 + (-1)^{N_z} \tilde {\mathcal C}_4 \mathcal M = 0.
\end{align}
From these relations and 
$\tilde {\mathcal C}_4^2 = - {\mathcal C}_4^2$,  
there are Kramers pairs if the states have the real eigenvalue ($\pm 1$) of $\mathcal C_4$, for
the even numbers of layers in each mirror sector.

\section{Topological number in the even numbers of layers}
\label{mz2}

The topological number $\zeta_{\rm M}$ Eq. (\ref{zeta}) is a $\mathbb Z$ topological invariant protected by translational, reflection, and time-reversal symmetries.
Definition and numerics for $\zeta_{\rm M}$ are shown based on Refs. \onlinecite{yu11, alexandradinata12}.

The (001)--films with odd numbers of layers has (001)--reflection symmetry.
This allows us to define the mirror Chern number since time-reversal symmetry is broken within each mirror sector.
Such a symmetry is called R$_-$ in Refs. \onlinecite{chiu13, morimoto13}.
The films with even numbers of layers, on the other hand, does not have (001)--reflection symmetry itself but they have the corresponding glide symmetry.
It is possible to decompose the Hamiltonian of the even numbers of layers into two independent sectors in the same way as in the odd numbers of layers (Appendix \ref{001}).
But the mirror Chern number in the even numbers of layers always vanishes since time-reversal symmetry arises within each mirror sector (Appendix \ref{trs}). 
This symmetry is called R$_+$.
For class AII--R$_+$ in two spatial dimensions, the possible topological invariant is the $\mathbb Z_2$ number, which is the same as that in two-dimensional TI with time-reversal symmetry.\cite{chiu13, morimoto13}
However, this is not the case for the present problem since
the Brillouin zone in each mirror sector is doubled (Appendix \ref{periodicity}) and 
time-reversal invariant momenta are located only at 
the $\bar{\rm X}$ points, which are inner points in the doubled Brillouin zone,
not at the zone center $\bar \Gamma$ point.
This type of time-reversal symmetry stemming from glide symmetry is different from that in other systems: time-reversal invariant momenta are located at the zone center and boundary in conventional cases.
Thus the $\mathbb Z_2$ invariant does not work in the even numbers of layers.
Therefore, we introduce another topological invariant consistent with glide and time-reversal symmetries in two spatial dimension, as follows.

\subsection{Definition}

Each mirror sector for the even numbers of layers have the doubled periodicity Eq. (\ref{period2}) in the momentum space.
The Brillouin zone is also doubled ($-2\pi \leq k_x, k_y \leq 2\pi$) as denoted by the solid lines in Fig. \ref{domain2}.
We define a topological number $\zeta_{\rm M}=(\zeta_+ - \zeta_-)/2$ in the even numbers of layers in the doubled Brillouin zone, where $\zeta_\pm$ 
 is defined by the winding number of the Wannier center;
\begin{align}
 \zeta_{\pm} &= 
\frac{1}{2\pi}
 \int_{0}^{2 \pi} dk_y 
\frac{\partial \theta_{\pm}(k_y)}{\partial k_y} 
- \frac{\theta_{\pm}(2 \pi) - \theta_{\pm}(0)}{2\pi} 
\in \mathbb Z.
\label{zetapm}
\end{align}
The Wannier center $\theta_{\pm}$ is obtained from the Wilson loop as
\begin{align}
 \theta_{\pm}(k_y) = \mathrm{Im} \ln \det D_\pm(k_y) = \mathrm{Im} \mathrm{Tr} \ln D_\pm(k_y),
\end{align}
with
\begin{align}
 D_\pm(k_y) = \mathrm P \exp
 \left[
  i \int_{-\pi}^{\pi} dk_x A_\pm(\bm k)
 \right]
 B_\pm(k_y).
\end{align}
The non-Abelian Berry connection $A_\pm(\bm k)$ and matrix $B_\pm(k_y)$ are given by
\begin{align}
 A_\pm(\bm k) = -i V^\dag_\pm(k_x, k_y) \frac{\partial V_\pm(k_x, k_y)}{\partial k_x},
\end{align}
and
\begin{align}
 B_\pm(k_y) = V^\dag_\pm(\pi, k_y) V_\pm(-\pi, k_y),
\end{align}
where matrix $V_\pm(k_x, k_y)$ is defined by
\begin{align}
 V_\pm(k_x, k_y) = \begin{pmatrix}
   | k_x, k_y, 1, \pm \rangle, \cdots, | k_x, k_y, 2N_\pm, \pm \rangle
 \end{pmatrix},
\end{align}
with $| k_x, k_y, m, \pm \rangle$, $m=1, \cdots, 2N_\pm$, an eigenvector of the Hamiltonian for occupied states.
Here $m$ is an occupied band index in each mirror sector, and $N_\pm$ is the number of the occupied states.

For numerical calculation, the following discretized form is useful,
\begin{align}
 \theta_\pm(k_y)
 = \sum_{i=0}^{N_x-1} \mathrm{Im} \ln \det F_{i,i+1, \pm}(k_y)
 + \mathrm{Im} \ln \det B_\pm(k_y),
\end{align}
where 
\begin{align}
 F_{i,i+1, \pm}(k_y) = V^\dag_\pm(k_i, k_y) V_\pm(k_{i+1}, k_y),
\end{align}
with $k_i = -\pi + 2\pi i/N_x$ and $i=0, \cdots, N_x$.
Note that, in the numerical calculation, the following must be explicitly taken into account to preserve gauge symmetry. 
$V^\dag_\pm(k_{i+1},k_y)$ in $F_{i+1, i+2, \pm}(k_y)$ is the hermitian conjugate of $V_\pm(k_{i+1}, k_y)$ in $F_{i,i+1, \pm}(k_y)$.
Similarly, $V^\dag_\pm(\pi,k_y)$ and $V_\pm(-\pi, k_y)$ in $B_\pm(k_y)$ are the hermitian conjugates of $V_\pm(\pi, k_y)$ in $F_{N_x-1, N_x}(k_y)$ and $V^\dag_\pm(-\pi, k_y)$ in $F_{0,1,\pm}(k_y)$, respectively.

The glide--winding number $\zeta_{\rm M}$ can be extended to three-dimensional systems.
If a system has glide symmetry as $H(k_x, k_y, 0) = M_0 H(k_x, k_y, 0) M_0^\dag$ and $H(k_x, k_y, \pi) = M_\pi H(k_x, k_y, \pi) M_\pi^\dag$ ($M_0 \ne M_\pi$ in general), 
a set of glide--winding numbers $\zeta_{\rm M}|_{k_z=0}$ and $\zeta_{\rm M}|_{k_z=\pi}$ at the symmetric planes of $k_z=0$ and $k_z=\pi$ are the $\mathbb Z \times \mathbb Z$ invariant in three spatial dimension.

\subsection{Symmetry}
 \label{symwn}

Quantity related to physical phenomena must be gauge invariant.
Here we show the topological number $\zeta_\pm$ is invariant under $\mathrm U(2N_\pm)$--gauge transformation.
Furthermore, we explain that
this number is quantized to an integer protected by spatial inversion symmetry.

\subsubsection{Gauge symmetry}

$\theta_\pm(k_y)$ is $\mathrm U(2N_\pm)$--gauge invariant, as shown below.
For 
$V_\pm(k_x, k_y) \to V_\pm(k_x, k_y) w_\pm(\bm k)$,
with $w_\pm(\bm k) \in \mathrm U(2N_\pm)$,
the non-Abelian Berry connection is transformed to
\begin{align}
 A_\pm(\bm k) \to w_\pm^\dag(\bm k) A_\pm(\bm k) w_\pm(\bm k)
 -i w_\pm^\dag(\bm k) \frac{\partial w_\pm(\bm k)}{\partial k_x},
\end{align}
and its integral of U(1) part is given by
\begin{align}
&
 \int_{-\pi}^\pi dk_x 
 \mathrm{Tr} A_\pm(\bm k)
 \nonumber \\ &
 \to 
 \int_{-\pi}^{\pi} dk_x
 \mathrm{Tr} A_\pm(\bm k)
 -i \int_{-\pi}^{\pi} dk_x \frac{\partial}{\partial k_x} \mathrm{Tr} \ln w_\pm(\bm k)
 \nonumber \\ &
 = 
 \int_{-\pi}^{\pi} dk_x
 \mathrm{Tr} A_\pm(\bm k)
 \nonumber \\ & \qquad
 -i \mathrm{Tr} \ln w_\pm(\pi, k_y) + i \mathrm{Tr} \ln w_\pm(-\pi, k_y),
\end{align}
And also, matrix $B_\pm(k_y)$ is transformed to
\begin{align}
 \mathrm{Tr} \ln B_\pm(k_y) 
&\to  
\mathrm{Tr} \ln B_\pm(k_y) 
\nonumber \\ & \hspace{-1em}
- \mathrm{Tr} \ln w_\pm(\pi, k_y) 
+
\mathrm{Tr} \ln w_\pm(-\pi, k_y).
\end{align}
These relations proves that the Wannier center $\theta_\pm(k_y)$ is gauge invariant.

\subsubsection{Inversion symmetry}

\begin{figure*}
\centering
\includegraphics[scale=0.9]{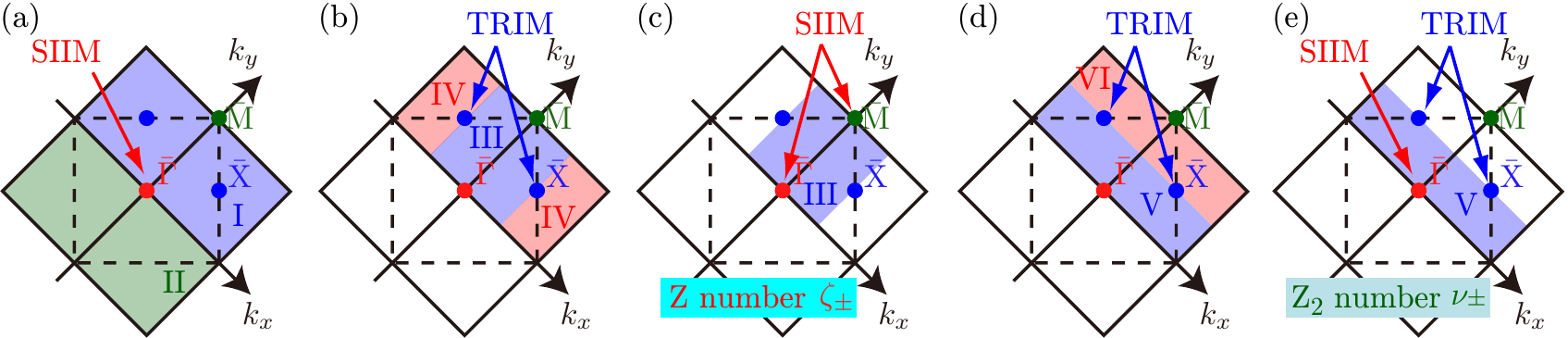}
\caption{
Domain of integration for topological numbers.
The dashed and solid lines indicate the projected Brillouin zone and
that in each mirror sector for the even numbers of layers, respectively.
The $\bar \Gamma$ $(k_x=0, k_y=0)$ and $\bar {\rm M}$ $(2\pi, 0)$ points are spatial-inversion invariant momenta (SIIM) and the $\bar {\rm X}$ point $(\pi, \pi)$ is a time-reversal invariant momentum (TRIM).
Regions I--VI denote the domain reduced by symmetries.
Topological numbers defined in regions III and V are $\mathbb Z$ ($\zeta_\pm$) and $\mathbb Z_2$ ($\nu_\pm$) numbers, respectively.
}
\label{domain2}
\end{figure*}

Inversion symmetry is preserved in both the even and odd numbers of layers (see Appendix \ref{inv}), i.e., inversion operator $\mathcal P$ can be restricted onto the space spanned by the $2N_\pm$ occupied states; $\mathcal P \to \mathcal P_\pm \equiv P_\pm \mathcal P P_\pm \in \mathrm U(2N_\pm)$.
The eigenvectors are transformed to
\begin{align}
 V_\pm(k_x, k_y) \to V_\pm(-k_x, -k_y) \mathcal P_\pm.
\end{align}
Therefore, one obtains
\begin{align}
 A_\pm(\bm k)  
 &= - \mathcal P_\pm^\dag A_\pm(-\bm k) \mathcal P_\pm,
 \\
 B_\pm(k_y) &= \mathcal P_\pm^\dag B_\pm^\dag(-k_y) \mathcal P_\pm,
\end{align}
namely the Wannier center satisfies the following relation,
\begin{align}
\label{wsis}
 \theta_\pm(k_y) = -\theta_\pm(-k_y).
\end{align}
This means that a winding number defined in the entire Brillouin zone ($-2\pi < k_y < 2\pi$, region I $\cup$ II in Fig. \ref{domain2}) is twice as large as that in half the Brillouin zone ($0 < k_y < 2\pi$, region I in Fig. \ref{domain2}).
This is why the topological number $\zeta_\pm$ is defined in $0 < k_y < 2\pi$ in Eq. (\ref{zeta}).
Moreover, at the spatial-inversion invariant momenta $k_y=0$ and $k_y=2\pi$, the Wannier center is fixed to $\theta_\pm(0), \theta_\pm(2\pi) = 0$ or $\pi$, which removes $2 n \pi$--ambiguity in the winding number.
As a result, the winding number $\zeta_\pm$ is proven to be a $\mathbb Z$ topological invariant.

\subsubsection{Time-reversal symmetry}

Time-reversal symmetry Eq. (\ref{trs1}) is preserved in the system but not in each mirror sector: $\hat \Theta$ interchanges the mirror-even and mirror-odd sectors as
\begin{align}
 A_\pm(\bm k) &= \mathcal T A_\mp^{*}(-\bm k) \mathcal T^{-1},
 \\
 B_\pm(k_y) &= \mathcal T B^{\rm T}_\mp(-k_y) \mathcal T^{-1},
\end{align}
with $\mathcal T$ being a skew matrix.
As a result, the Wilson loop satisfies
\begin{align}
 D_\pm(k_y) = \mathcal T D_\mp^{\rm T}(-k_y) \mathcal T^{-1}.
\end{align}
which implies that the Wannier centers in each mirror sector are not independent,
\begin{align}
\label{wtrs}
 \theta_\pm(k_y) = \theta_{\mp}(-k_y).
\end{align}
Equations (\ref{wsis}) and (\ref{wtrs}) result in $\zeta_+ + \zeta_- = 0$.
Only the difference $\zeta_{\rm M} = (\zeta_+ - \zeta_-)/2$ can take a finite value.

The even numbers of layers have the additional time-reversal symmetry
at the $\rm \bar X$ point (see Appendix \ref{trs}).
The non-Abelian Berry connection around the $\rm \bar X$ point ($= \bm G_i/2$) is transformed  as
\begin{align}
\label{trsx}
 A_\pm(\bm k + \bm G_i/2) 
=
\mathcal T_{\rm \bar X} 
A_\pm^{*}(-\bm k + \bm G_i/2) 
\mathcal T_{\rm \bar X}^{-1}.
\end{align}
Suppose the Wilson loop $D'_\pm(k_y)$ defined in region I [Fig. \ref{domain2}(a)] as
\begin{align}
 D'_\pm(k_y) = \mathrm P \exp
 \left[
   i \int_{-2\pi}^{2\pi} dk_x A_\pm(\bm k)
 \right],
\end{align}
where $V_\pm(2\pi,k_y) = V_\pm(-2\pi, k_y)$ is fixed due to Eq. (\ref{period2}) $H_\pm(\bm k) = H_\pm(\bm k + \bm G_1 + \bm G_2)$.
The Wannier center $\theta'_\pm(k_y) = \mathrm{Im} \ln \det D'_\pm(k_y)$ satisfies the following relation,
\begin{align}
 \theta'_\pm(k_y+\pi) 
= \theta'_\pm(-k_y+\pi),
\end{align}
since $D'_\pm(k_y+\pi) = \mathcal T_{\rm \bar X} D'^{\rm T}_\pm(-k_y + \pi) \mathcal T_{\rm \bar X}^{-1}$.
Namely, a topological number defined in entire region I ($-2\pi \leq k_x \leq 2\pi, 0 \leq k_y \leq 2\pi$) in Fig. \ref{domain2}(a) vanishes.
In order to extract nontrivial contribution of topological number, the domain is divided so that one half of Kramers pair is picked up.
There are two ways to divide region I.
One is I $\to$  III + IV [Fig. \ref{domain2}(b)].
That defined in region III [Fig. \ref{domain2}(c)] is the $\mathbb Z$ topological invariant, which is already introduced in Eq. (\ref{zetapm}).
The other is I $\to$ V + VI [Fig. \ref{domain2}(d)].
A topological number $\nu_\pm$ defined in region V [Fig. \ref{domain2}(e)],
\begin{align}
 \nu_\pm = \int_0^{\pi} \frac{dk_y}{2\pi}
 \frac{\partial \theta'_\pm(k_y)}{\partial k_y}
 -
 \frac{\theta'_\pm(\pi) - \theta'_\pm(0)}{2\pi},
\end{align}
 is a $\mathbb Z_2$ topological invariant, since eigenvalues of $D'_\pm(\pi)$ are doubly degenerated due to $\mathcal T_{\rm \bar X} = -\mathcal T^{\rm T}_{\rm \bar X}$.
The two-fold degeneracy leads to $4 n \pi$ (not $2 n \pi$) phase ambiguity in $\theta'(\pi)$.
The spatial inversion symmetry requires $\theta'(0)=0$, as in Eq. (\ref{wsis}).
In consequence, $\nu_\pm \mod 2$ is a $\mathbb Z_2$ topological invariant.
In the actual calculation, however, $\nu_\pm = 0 \mod 2$ is obtained.
A nontrivial topological number in the even numbers of layers comes from $\zeta_{\rm M}$, which is explained in the following section.

\subsection{Wannier-center flow}

\begin{figure}
\centering
\includegraphics{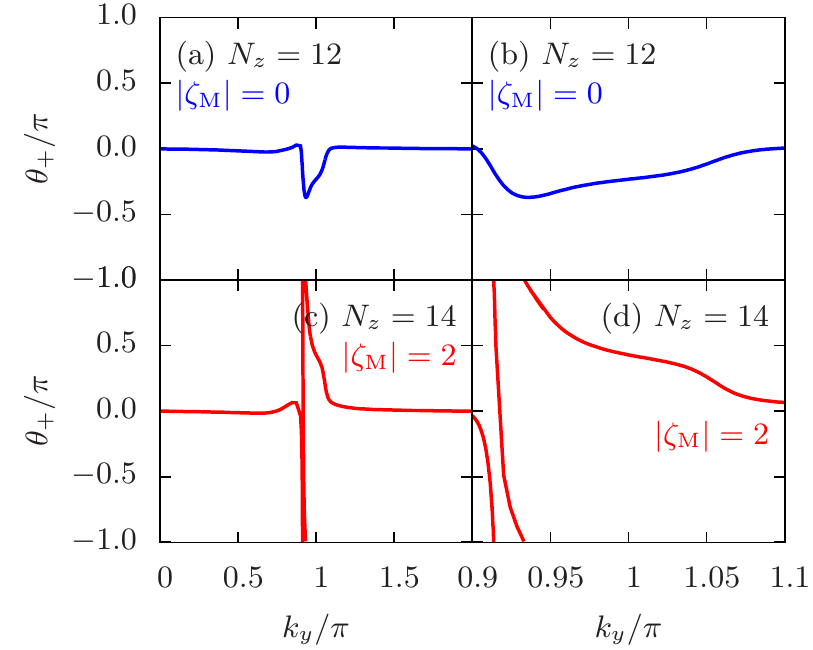}
\caption{Evolution of Wannier center in the mirror-even sector of SnTe films for $N_z=12$ ($\zeta_{\rm M}=0$) [(a) and (b)] and $N_z=14$ ($|\zeta_{\rm M}|=2$) [(c) and (d)].
}
\label{12-14}
\end{figure}

Here we show examples of the Wannier-center flow and the corresponding topological number $\zeta_{\rm M}$ denoted in Fig. \ref{pd_even}(a).
Figure \ref{12-14} shows the Wannier-center flow around the transition from $\zeta_{\rm M} = 0$ to $|\zeta_{\rm M}|=2$.
For $N_z=12$, the Wannier center $\theta_+$ in the mirror-even sector does not wind [Figs. \ref{12-14}(a) and \ref{12-14}(b)].
For $N_z=14$, on the other hand, the Wannier center goes through the branch cut ($\theta = \pm \pi$) twice at $k \sim 0.914 \pi$ and $k \sim 0.933 \pi$.
This results in the winding number of $\theta_+$ is two, i.e., the corresponding topological number is given by $|\zeta_{\rm M}|=2$.

\begin{figure}
\centering
\includegraphics{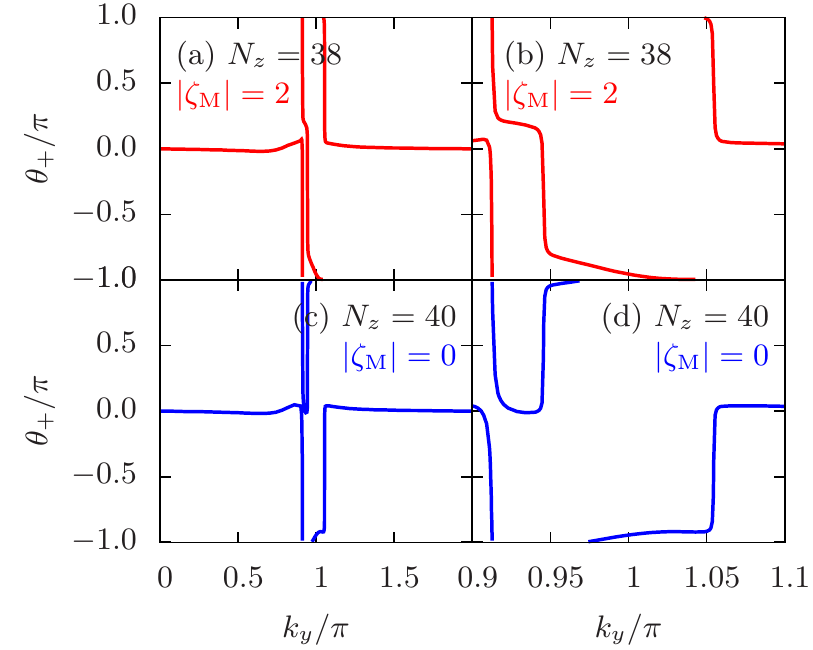}
\caption{Evolution of Wannier center in the mirror-even sector of SnTe films for $N_z=38$ ($|\zeta_{\rm M}|=2$) [(a) and (b)] and $N_z=40$ ($\zeta_{\rm M}=0$) [(c) and (d)].
}
\label{38-40}
\end{figure}

Figure \ref{38-40} shows the Wannier-center flow for $N_z=38$ and $N_z=40$, where the topological number changes from $|\zeta_{\rm M}|=2$ to $\zeta_{\rm M}=0$.
For $N_z=38$, the Wannier center goes through the branch cut at $k_y \sim 0.913\pi$ and $k_y \sim 1.045 \pi$, i.e., the winding number is obtained to be $|\zeta_{\rm M}|=2$.
On the other hand, for $N_z=40$, the Wannier center goes through the branch cut from above  at $k_y \sim 0.913 \pi$, but from below at $k_y \sim 0.97 \pi$.
Hence the winding number vanishes; $\zeta_{\rm M}=-1+1=0$ for $N_z=40$.

\bibliography{note}

\clearpage

\end{document}